\documentclass[traditabstract]{aa} 

\usepackage{graphicx}
\usepackage{txfonts}
\usepackage{natbib}
\bibpunct{(}{)}{;}{a}{}{,}
\usepackage{float}
\usepackage{calc}
\usepackage{textcomp}
\usepackage{placeins}
\usepackage{color}
\usepackage{rotating} 
\usepackage{lscape}
\usepackage{url}
\usepackage[colorlinks=true,linkcolor=blue,citecolor=blue]{hyperref}
\usepackage{subfig}
\usepackage[all]{draftcopy}
\usepackage{longtable}
\usepackage{array}
\usepackage{threeparttable}
\usepackage{lscape}
\DeclareTextSymbol{\degre}{OT1}{23}
\newcounter{savedfootnote}

\def \microns{{\,$\mu$m}}
\begin{document}
   \title{A hyper luminous starburst at $z=4.72$ magnified by a lensing galaxy pair at $z=1.48$.}

\author{L.~Ciesla\inst{1,2},
	 M.~B\'ethermin\inst{1},
	 E.~Daddi\inst{2}, 
	 J.~Richard\inst{3},
	 T.~Diaz-Santos\inst{4},
	 M.~T.~Sargent\inst{5},
	 D.~Elbaz\inst{2},
	 M.~Boquien\inst{6},
	 T.~Wang\inst{7},
	 C.~Schreiber\inst{8},
	 C.~Yang\inst{9},
	 J.~Zabl\inst{3},
	 M.~Fraser\inst{10},
	 M.~Aravena\inst{4}
	 R.~J.~Assef\inst{4},
	 A.~J.~Baker\inst{11},
	 A.~Beelen\inst{1},
	 A.~Boselli\inst{1},
	 F.~Bournaud\inst{2},
	 D.~Burgarella\inst{1},
	 V.~Charmandaris\inst{12,13},
	 P.~C\^ot\'e\inst{14},
	 B.~Epinat\inst{1},
	 L.~Ferrarese\inst{14},
	 R.~Gobat\inst{15},
	 and O.~Ilbert\inst{1}.
    }

\institute{	
Aix Marseille Univ, CNRS, CNES, LAM, Marseille, France
\and
Laboratoire AIM-Paris-Saclay, CEA/DSM/Irfu - CNRS - Universit\'e Paris Diderot, CEA-Saclay, F-91191 Gif-sur-Yvette, France	
\and
Univ Lyon, Univ Lyon1, Ens de Lyon, CNRS, Centre de Recherche Astrophysique de Lyon UMR5574, F-69230 Saint-Genis-Laval, France
\and
Nucleo de Astronomia, Facultad de Ingeniería y Ciencias, Universidad Diego Portales, Ejercito Libertador 441, Santiago, 8320000, Chile
\and
Astronomy Centre, Department of Physics and Astronomy, University of Sussex, Brighton, BN1 9QH, UK
\and
Centro de Astronomía (CITEVA), Universidad de Antofagasta, Avenida Angamos 601, Antofagasta, Chile
\and
Institute of Astronomy, Graduate School of Science, The University of Tokyo, 2-21-1 Osawa, Mitaka, Tokyo 181-0015, Japan
\and
Department of Physics, University of Oxford, Keble Road, Oxford, OX1 3RH, UK
\and
European Southern Observatory, Alonso de Córdova 3107, Casilla, 19001, Vitacura, Santiago, Chile
\and
School of Physics, O'Brien Centre for Science North, University College Dublin, Belfield, Dublin 4, Ireland
\and
Department of Physics and Astronomy, Rutgers, the State University of New Jersey, 136 Frelinghuysen Road, Piscataway, NJ 08854-8019, USA
\and
Department of Physics, University of Crete, GR-71003 Heraklion, Greece
\and
Institute of Astrophysics, Foundation for Research and Technology-Hellas, GR-71110 Heraklion, Greece
\and
National Research Council, 5071 W. Saanich Road, Victoria, BC V9E 2E7, Canada
\and
Instituto de Fisica, Pontificia Universidad Cat\'olica de Valpara\'iso, Casilla 4059, Valparaiso, Chile
}

   \date{Received; accepted}

  \abstract
{
We serendipitously discovered in the \textit{Herschel} Reference Survey an extremely bright infrared source with $S_{500}\sim120$\,mJy in the line of sight of the Virgo cluster which we name Red Virgo 4 (RV4).
Based on IRAM/EMIR and IRAM/NOEMA detections of the CO(5-4), CO(4-3), and [CI] lines, RV4 is located at a redshift of 4.724, yielding a total observed infrared luminosity of 1.1$\pm$0.6$\times$10$^{14}$\,L$_{\odot}$.
At the position of the \textit{Herschel} emission, three blobs are detected with the VLA at 10\,cm. 
The CO(5-4) line detection of each blob confirms that they are at the same redshift with the same line width, indicating that they are multiple images of the same source.
In \textit{Spitzer} and deep optical observations, two sources, High-z Lens 1 (HL1) West and HL1 East, are detected at the center of the three VLA/NOEMA blobs. 
These two sources are placed at $z=1.48$ with XSHOOTER spectra, suggesting that they could be merging and gravitationally lensing the emission of RV4.
HL1 is the second most distant lens known to date in strong lensing systems.
Constrained by the position of the three VLA/NOEMA blobs, the Einstein radius of the lensing system is 2.2\arcsec$\pm$0.2 (20\,kpc). 
The high redshift of HL1 and the large Einstein radius are highly unusual for a strong lensing system.
In this paper, we present the insterstellar medium properties of the background source RV4.
Different estimates of the gas depletion time yield low values suggesting that RV4 is a starburst galaxy.
Among all high-z submillimeter galaxies (SMGs), this source exhibits one of the lowest L$_{[CI]}$ to L$_{IR}$ ratios, 3.2$\pm$0.9$\times$10$^{-6}$, suggesting an extremely short gas depletion time of only 14$\pm$5 Myr. 
It also shows a relatively high L$_{[CI]}$ to L$_{CO(4-3)}$ ratio (0.7$\pm$0.2) and low L$_{CO(5-4)}$ to L$_{IR}$ ratio (only $\sim$50$\%$ of the value expected for normal galaxies) hinting a low density of gas. 
Finally, we discuss the short depletion time of RV4. It can be explained by either a very high star formation efficiency, which is difficult to reconcile with major mergers simulations of high-$z$ galaxies, or a rapid decrease of star formation, which would bias the estimate of the depletion time toward artificially low value.
}

  
   \authorrunning{Ciesla et al.}
   \titlerunning{A $z=4.72$ starburst lensed by a merging pair of $z=1.48$ galaxies.}

   \maketitle

%=================================================================================
\section{Introduction}
In the local Universe, a third of the total bolometric luminosity of galaxies is emitted in the infrared (IR) and submillimetric (submm) domains by dust grains, which reprocess  the energy absorbed from the stars and active galactic nuclei (AGN) \citep{SoiferNeugebauer91}.
In submm galaxies \citep[SMGs,][]{Smail97,Barger98}, highly dust-obscured systems, the energy balance between optical and IR/submm domains is skewed even more towards long-wavelength emission.
The dust is heated by numerous young stars causing the SMGs to be extremely bright in the submm regime and reach huge IR luminosities (L$_{IR}$) higher than $\sim$ 10$^{12}$\,L$_{\odot}$. 	
More and more of these extreme star-forming galaxies are being found \citep[e.g.,][]{Daddi09,Negrello10,Frayer11,Yun12,Riechers13,Weiss13,Ivison13,Vieira13,Dowell14,Canameras15,DiazSantos16,Strandet17,Marrone18,Zavala18}, among which are sources at $z>$ 4, indicating an extremely rapid assembly of these structures.
Their high SFRs suggest the presence of large gas reservoirs \citep[e.g., a gas fraction of $\sim$50\%;][]{Fu13, Bethermin15,Aravena16} and their large dust content indicates that their interstellar medium (ISM) is metal enriched. 
One of the challenges posed by these sources is to understand how this massive and mature ISM can be in place so early in these galaxies.
Indeed, these massive galaxies are still not well reproduced by cosmological simulations \citep{Dave10,Cousin15,Sparre15}.
The study of these starbursting systems, which usually lie above the main-sequence of star-forming galaxies \citep[e.g.,][]{Noeske07_SFseq,Elbaz07,Elbaz11} and whose mid-IR properties resemble those of the very brightest, nearby IR-luminous galaxies \citep{DiazSantos11}, is of paramount importance for providing constraints on models of the formation and evolution of massive galaxies. 
In fact, the massive ellipticals that we observe in the local Universe could have formed rapidly $\sim\,$10\,Gyr ago and be the remnant of this population of starbursting galaxies at high redshift \citep{McCarthy04,Daddi07a,Daddi07b,Tacconi08,Cimatti08,Haan13}. 
Within this scenario, because of their tremendous SFRs, the most vigorously star-forming SMGs rapidly quench their star formation by exhausting their gas reservoir in only a few hundred Myr, making them very rare.
For instance, \cite{Fu13} estimate the space density of SMGs with SFR $>$ 2000\,M$_{\odot}$ yr$^{-1}$ to be 1.4$\times$ 10$^{-5}$\,Mpc$^{-3}$.

Nevertheless, samples of SMGs have been built with facilities like the South Pole Telescope (SPT) and the \textit{Herschel} Space Observatory \citep[e.g.,][]{Eales10HA,Negrello10,Bussmann13,Weiss13,Vieira13}, and statistical studies of their properties can be now found in the literature.
Studies of high-$z$ bright SMGs (S$_{850}>$50\,mJy) show that the majority of these sources are gravitationally lensed by  foreground, massive galaxies, thus amplifying the background source emission. 
This magnification allows us to detect high-redshift galaxies that could have been otherwise missed or observed with a lower signal-to-noise ratio.
The ISM of high-$z$ sources can thus be studied through the easier detection of lines, thanks to flux boosting \citep[e.g.,][]{Weiss13,Alaghband13,Bethermin15,Bothwell17,Yang17,Cunningham19}.
Another aspect of gravitational lensing is that the deflection of the light emitted by a background source allows us to probe the mass distribution of the foreground source acting as the lens, constraining dark matter (DM) sub-halo structures \citep{Hezaveh16} and/or the initial mass function \citep[IMF;][]{Canameras17}, for instance.
Finally, gravitational lensing boosts the angular resolution with which we can observe background sources allowing spatially resolved studies of the ISM of high redshift galaxies \citep[e.g.,][]{Swinbank15,Canameras18,Litke19,YangC19,Apostolovski19}.

In this study, we report the discovery of a peculiar strong lensing system.
The high-z galaxy, HRS188.6868+7.1357, hereafter Red Virgo 4 (RV4), was first detected in a survey of local galaxies carried out with \textit{Herschel}/SPIRE \citep{Griffin10,Boselli10a}.
It is detected in the background of the SPIRE pointing of IC\,3521 at RA: 188.6868 and Dec: +7.1357 (J2000).
The high flux density (122\,mJy at 500\microns) as well as the redward increasing spectrum led to interest in this source.
We gathered and obtained ancillary data to constrain its infrared (IR) emission, measure its redshift, identify the foreground source acting as a lens (High-z Lens 1, HL1, which turned to be one of the most distant lenses found until now) and constrain the strong lensing system.

In Sect.~\ref{data}, we list the set of ancillary data spanning from optical to radio, with both spectroscopic and photometric observations. 
The analysis of the data resulting in the lensing configuration conclusion is presented in Sect.~\ref{lensconf}.
The depletion time in particular is discussed in Sect.~\ref{shortdepl}.
The characterization of RV4 is developed in Sect.~\ref{rv4}, and conclusions are given in Sect.~\ref{conclusion}.
A companion paper focusing on a detailed lens model and on the nature of HL1 is in preparation.

Throughout the paper, we use the $\Lambda$CDM cosmology of \cite{Planck16} and a \cite{Salpeter55} IMF.
 
%=================================================================================
\section{\label{data}Data} 
In this section, we describe the set of data that we obtained for the characterization of the lensing system composed of RV4 and HL1.

    \subsection{Unresolved observations of RV4}
    \subsubsection{\textit{Herschel}}

    RV4 is detected in the background of the SPIRE (Spectral and Photometric Imaging Receiver)  images of IC3521, a Virgo cluster galaxy observed as part of the \textit{Herschel} Reference Survey \citep[HRS,][]{Boselli10a} and the \textit{Herschel} Virgo Cluster Survey \citep{Davies10}.
    As shown in Fig.~\ref{imagessub}, RV4 appears as a particularly bright, red source in SPIRE imaging of IC3521, clearly contrasting with the foreground galaxy.

    The FWHMs of the 250, 350, and 500\microns\ maps of IC3521 are 18.2\arcsec, 24.5\arcsec, and 36.0\arcsec, respectively \citep{Ciesla12}.
    RV4 is not resolved in these SPIRE images, and flux densities are extracted using the timeline-based PSF fitting approach which is the most appropriate method for point-like \textit{Herschel} sources \citep{Bendo13}.
    RV4's SPIRE flux densities are provided in Table~\ref{phot}.
    We refer the reader to \cite{Smith12} for a complete description of the SPIRE data reduction made as part of the HRS.
    
    IC3521 has also been observed with PACS \citep[Photodetector Array Camera and Spectrometer;][]{Poglitsch10} at 100 and 160\microns\ \citep{Cortese14}.
    RV4 is not detected in the PACS images, and the derived upper limits (see Table~\ref{phot}) do not provide any useful constraint on the IR spectral energy distribution (SED).
    
    \subsubsection{IRAM 30\,m/NIKA}
    After its initial detection by \textit{Herschel}, we used the NIKA camera \citep[Neel-IRAM-KID-Array;][]{Monfardini10} on the 30\,m telescope at Pico de Valeta (Spain) to follow up RV4 (234-14, PI: M.~B\'ethermin) at 1.2 and 2\,mm. 
    Nine Lissajous scans of 5 minutes were performed to detect the source, and the data were reduced using the standard pipeline. 
    At the IRAM 30m telescope's resolution at these wavelengths ($\sim$12\arcsec\ and 18\arcsec\ at 1.2 and 2\,mm, respectively), the RV4 system is unresolved; therefore a standard PSF-fitting extraction is used.
    Flux densities are provided in Table~\ref{phot}.

    \subsubsection{GBT/Zpectrometer}
    We obtained 6\,h of GBT\footnote{The Green Bank Telescope is a facility of the National Science Foundation operated under cooperative agreement by Associated Universities, Inc.}/Zpectrometer observations of RV4 (GBT14A-162, PI: L.~Ciesla) to measure its spectroscopic redshift.
    The observations were carried out on March 1$^{st}$, 2014.
    We aimed at the detection of the CO(1-0) line in the frequency range of the Zpectrometer (25.6-36.1\,GHz), covering $2.1<z<3.5$ with a spectral resolution of 32\,MHz per channel.
    No line was detected in the observations.
    Nevertheless, these observations allowed us to narrow the range of redshift solutions for RV4, and thus pinpoint the true redshift of the source.
    
    \subsubsection{IRAM 30m/EMIR}
    We obtained 6\,h of IRAM 30\,m telescope DDT time (D07-15, PI: L.~Ciesla) aiming at the detection of CO(5-4) and CO(4-3) to measure the redshift of RV4. 
    The observations were made on February, 24$^{th}$ and March, 1$^{st}$ of 2016 with EMIR in band E090 (3\,mm) over the 80-111\,GHz frequency range with a spectral resolution $\Delta v$ of 50\,km/s. 
    The wide-band line multiple auto-correlator (WILMA) and the fast Fourier Transform Spectrometer (FTS-200) were used simultaneously as backends during the observations. 
    Bright planet/quasar calibrators including Jupiter and J1226+023 were used for pointing and focusing. The weather conditions were excellent with $\tau_\mathrm{225\,GHz} \lesssim 0.2$, reaching a sensitivity of $\lesssim$\,0.6\,mK per 50\,km/s and a system temperature of $\lesssim 100$\,K.
    The data were calibrated using the standard dual method. Data were then reduced with the \texttt{GILDAS}\footnote{See \url{http://www.iram.fr/IRAMFR/GILDAS} for more information about the GILDAS software.} package \texttt{CLASS}.  
    The baseline-removed spectral scans were co-added according to the weights derived from the noise levels of each. 
    We also include $\sim$\,10\% absolute flux calibration uncertainty in our overall uncertainty.
    
    \subsection{High resolution observations of RV4}
    \subsubsection{VLA}
    To spatially resolve the emission of RV4, we obtained 9.5\,h (7.4\,h on source) of Karl G. Jansky Very Large Array\footnote{The National Radio Astronomy Observatory is a facility of the National Science Foundation operated under cooperative agreement by Associated Universities, Inc.} (VLA) time (VLA/2014-06-044, PI: L.~Ciesla) in A-configuration with 28 antennas at 3\,GHz (S-band, 2.0\,GHz-3.9\,GHz), reaching a sensitivity of 2\,$\mu$Jy/beam and a beam size of 0.65\arcsec. 
    The observations were carried out from August 3$^{rd}$ to August 22$^{nd}$ 2015.
    The data were calibrated by the observatory pipeline. We produced a continuum image using the CASA software \citep{McMullin07}. 
    We used all the channels to produce a continuum map and maximize the SNR. 
    We used Briggs weighting with robust=0.5 to achieve a good compromise between sensitivity (2.6\,$\mu$Jy/beam) and angular resolution (0.68\arcsec$\times$0.55\arcsec).
    The flux densities of the detections are provided in Table~\ref{phot}.
     
    \subsubsection{NOEMA}
    We obtained 6\,h (3.1\,h on source) of PolyFiX NOEMA data (W17EG002, PI: L.~Ciesla) to map the CO(5-4) emission at observed-frame 100.6\,GHz, in the 3\,mm band (A-configuration, 9 antennas, 82.9-90.6\,GHz and 98.4-106.1\,GHz), with a native resolution of 2\,MHz, that we later rebinned by a factor of 8 before imaging.
    Observations were carried out on February 6$^{th}$ and 9$^{th}$, 2017. 
    We reached a sensitivity of 13\,$\mu$Jy/beam with a spectral resolution of 167.8\,MHz per channel, and  a spatial resolution of 1.56\arcsec$\times$0.84\arcsec.
    The data were calibrated using the \texttt{GILDAS/CLIC} package. 
    The data cubes and continuum maps were generated using \texttt{GILDAS/MAPPING}. 
    Because of the very large band width of Polyfix, we imaged separately the continuum in the lower and upper side bands.
    Both CO(5-4) and [CI] (492.161\,GHz rest frame) lines are clearly detected (see Sect.~\ref{secrv4}).
    
    \subsection{Observations of HL1}
    \subsubsection{\textit{Spitzer}}
    
    Mid-IR data from \textit{Spitzer}/IRAC (InfraRed Array Camera) are available only at 3.6 and 4.5\,$\mu$m, acquired as part of the The \textit{Spitzer} Survey of Stellar Structure in Galaxies \citep[S4G;][]{Sheth10}, and were downloaded from the NASA/IPAC Infrared Science Archive\footnote{\url{https://irsa.ipac.caltech.edu/data/SPITZER/S4G/}}. 
    We refer the reader to \cite{Sheth10}, \cite{MunozMateos13}, and \cite{Querejeta15} for detailed descriptions of the data acquisition and reduction.
    No observation with IRAC3 and 4, nor with MIPS, is available at the coordinates of RV4.

    \subsubsection{CFHT data from NGVS}
    Deep ground-based optical images of the Virgo cluster are available as part of the Next Generation Virgo cluster Survey \citep[NGVS;][]{Ferrarese12} obtained with MegaPrime (CFHT).
    The region around RV4 has been observed in $u^*$, $g'$, $i'$, and $z'$ bands.
    The depths for a point source with SNR of 10 are 25.9, 25.7, 24.9, and 24.6\,AB\,mag, in the $u^*$, $g'$, $i'$, and $z'$ bands, respectively.
    We refer the reader to \cite{Ferrarese12} and \cite{Licitra16} for detailed information on the data acquisition and reduction.

    \subsubsection{Magellan/FOURSTAR}
    We observed HL1 in March 2018 with the near-infrared imager Fourstar on the 6.5m Magellan Baade telescope using a random-position dither pattern. 
    Three filters were used: J1 (corresponding to the Y band), J, and Ks, with integrations of 44.8, 38.4, and 15.1\,min, respectively. 
    These data were initially calibrated with the FSRED pipeline.
    
    \subsubsection{VLT/XSHOOTER}
    We obtained 6\,h of VLT/XSHOOTER \citep{Vernet11} time as part of cycle 97A (097.A-0511, PI: T. Diaz-Santos) aiming to achieve an SNR of 5 in the 1.5-2.2\microns\ range (NIR arm).
    The observations were carried out on February 1$^{st}$ to 7$^{th}$, 2017.
    The total effective integration time of \(4.8\,\mathrm{h}\) and \(4.5\,\mathrm{h}\) in NIR and VIS arms, respectively, was split between 6 OBs with 4 nodding positions (ABBA) each. 
    Due to a telescope pointing issue, we needed to reject one of the 6 OBs. 
    We reduced each of the remaining 20 nodding pairs (AB) separately using the XSHOOTER pipeline \citep{Modigliani10}. 
    We flux-calibrated the data using standard pipeline recipes applied to observations of flux standard stars taken during each night of the observations. 
    In addition, we corrected the data for telluric absorption using a model atmospheric transmittance spectrum created with \textsc{molecfit} \citep{Smette15,Kausch15} from observations of telluric standard stars taken close in time and airmass to the science observations. 
    Subsequently, we optimally combined the individual nodding pairs (2D spectra) with a weighted average using our scripts. 
    Our scripts also corrected the wavelength scale to vacuum and removed the heliocentric velocity. 
    Finally, we extracted 1D spectra from the resulting 2D master spectrum.
    The 1.2\arcsec\ slits provide a nominal spectral resolution of R=6500 and R=4300 in VIS and NIR, respectively.
    
%=================================================================================
\section{\label{lensconf}An unusual strong lensing system} 

\subsection{\label{secrv4}Red Virgo 4: a magnified SMG}
\begin{figure}%[!h] 
  	\includegraphics[width=\columnwidth]{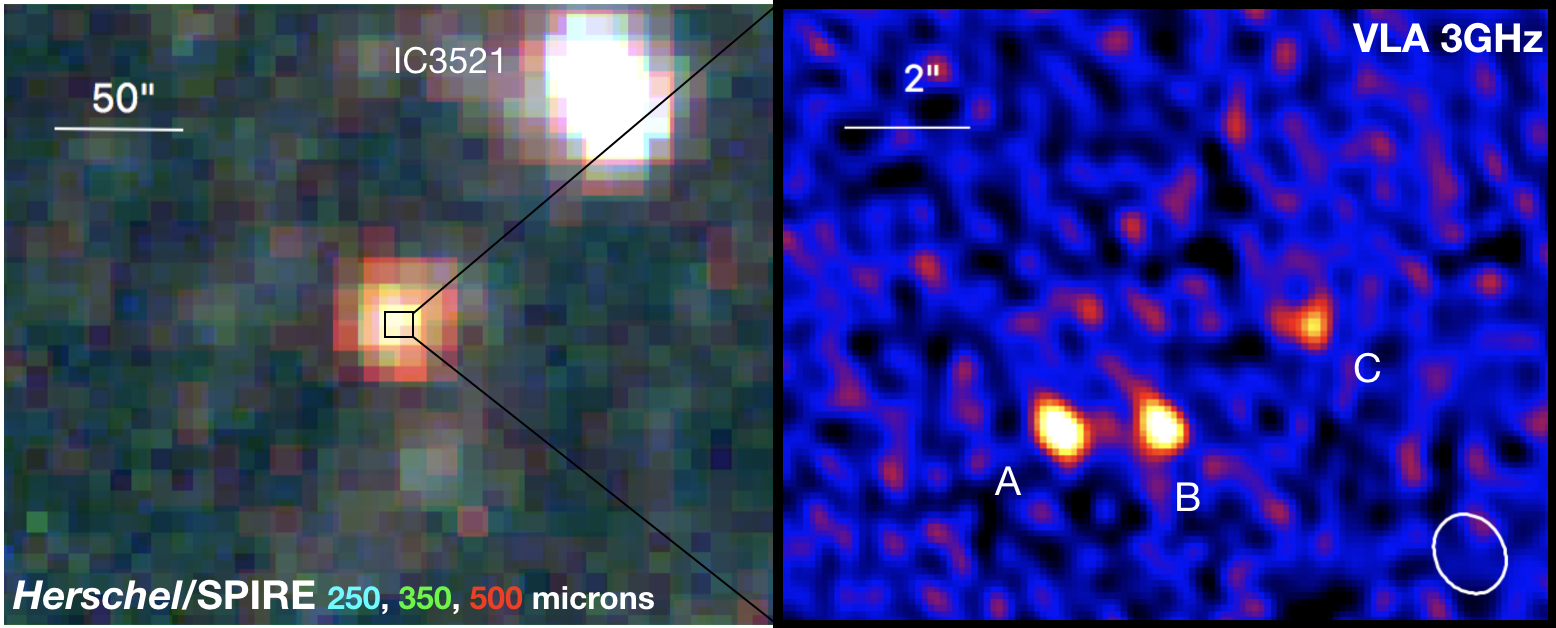}
  	\caption{ \label{imagessub} \textbf{Left panel:} \textit{Herschel}/SPIRE RGB color image (red: 500\microns; green: 350\microns; blue: 250\microns). \textbf{Right panel:} VLA 3\,GHz detections of RV4. The SPIRE emission splits into three unresolved sources at 3\,GHz. The beam size is indicated by the white ellipse.
	}
\end{figure}

\begin{figure}%[!h] 
  	\includegraphics[width=\columnwidth]{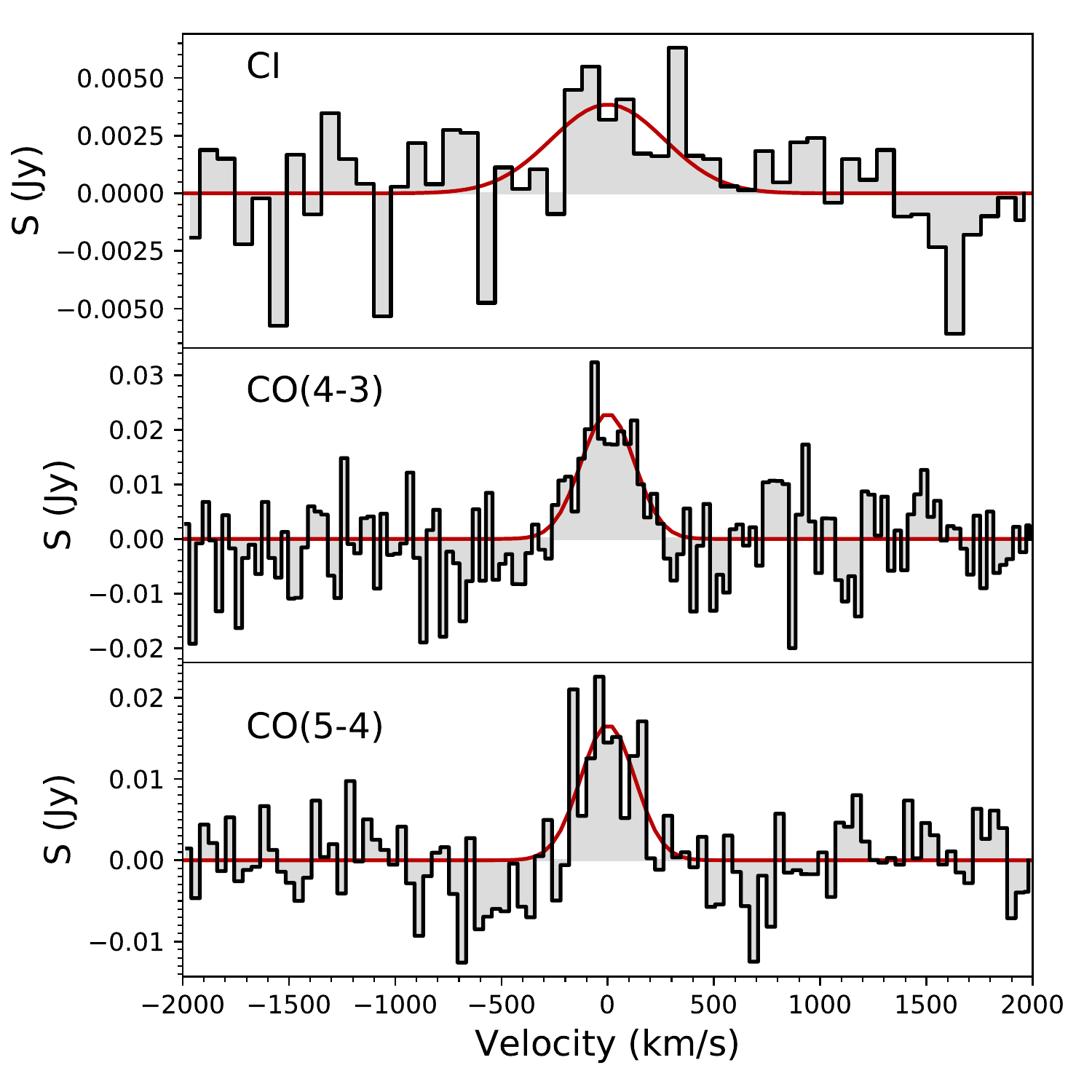}
  	\caption{ \label{emir} [CI](1-0), CO(4-3), and CO(5-4) emission lines detected with 30m/EMIR at the RV4 position. They correspond to the integrated emission of the three VLA/NOEMA blobs. The red solid lines shows the best fit of Gaussian. 
	}
\end{figure}

\begin{figure}%[!h] 
  	\includegraphics[width=\columnwidth]{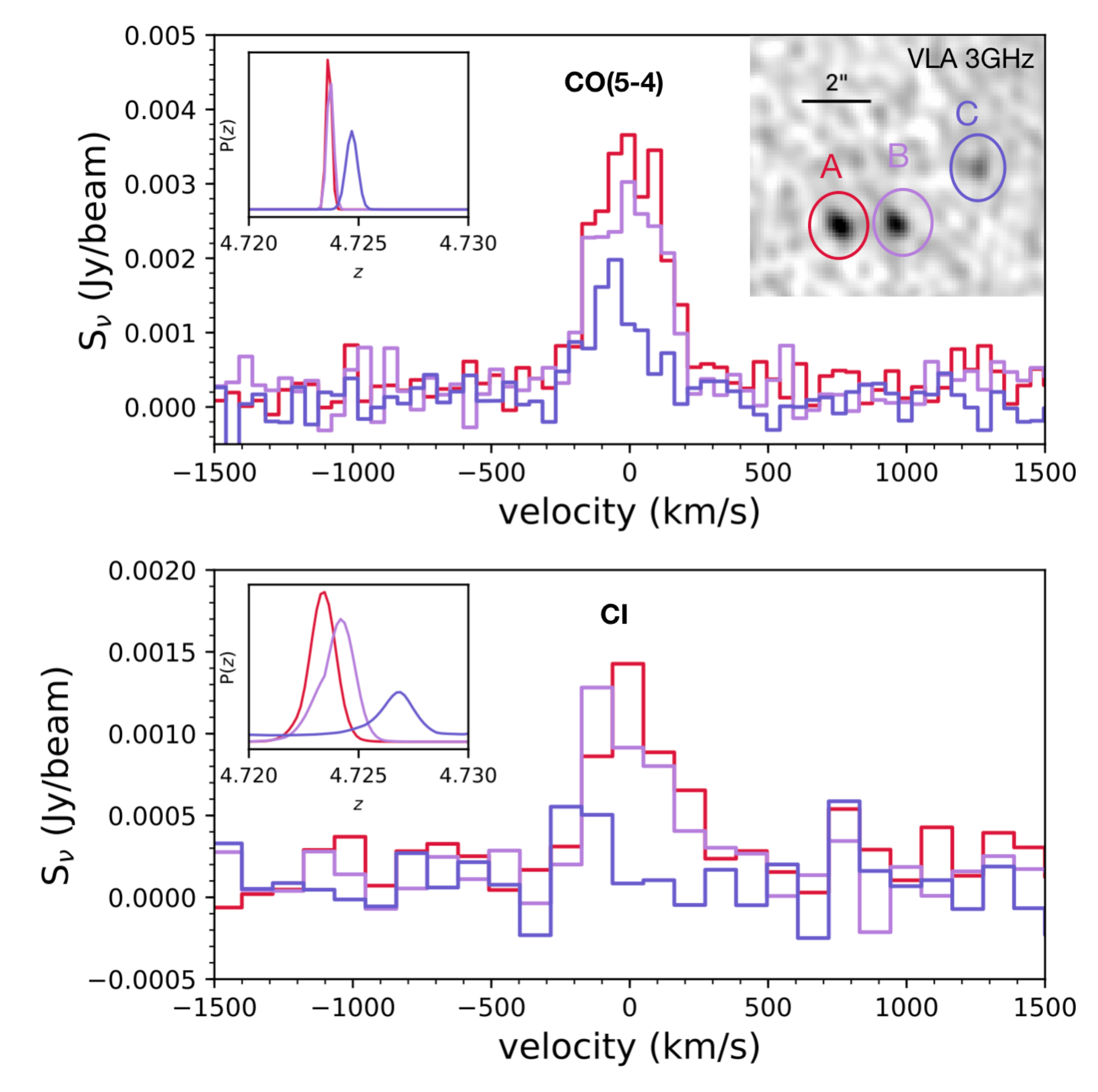}
  	\caption{ \label{noemalines} \textbf{Top panel:} NOEMA observations of CO(5-4) lines of blob A (red), B (purple), and C (blue). Inset panel: VLA 3\,GHz image of RV4. \textbf{Bottom panel:} [CI](1-0) lines of blob A (red), B (purple), and C (blue). The inset panels show the redshift probability distribution functions obtained with \textsc{slinefit}. 
	}
\end{figure}

\begin{figure}%[!h] 
  	\includegraphics[width=\columnwidth]{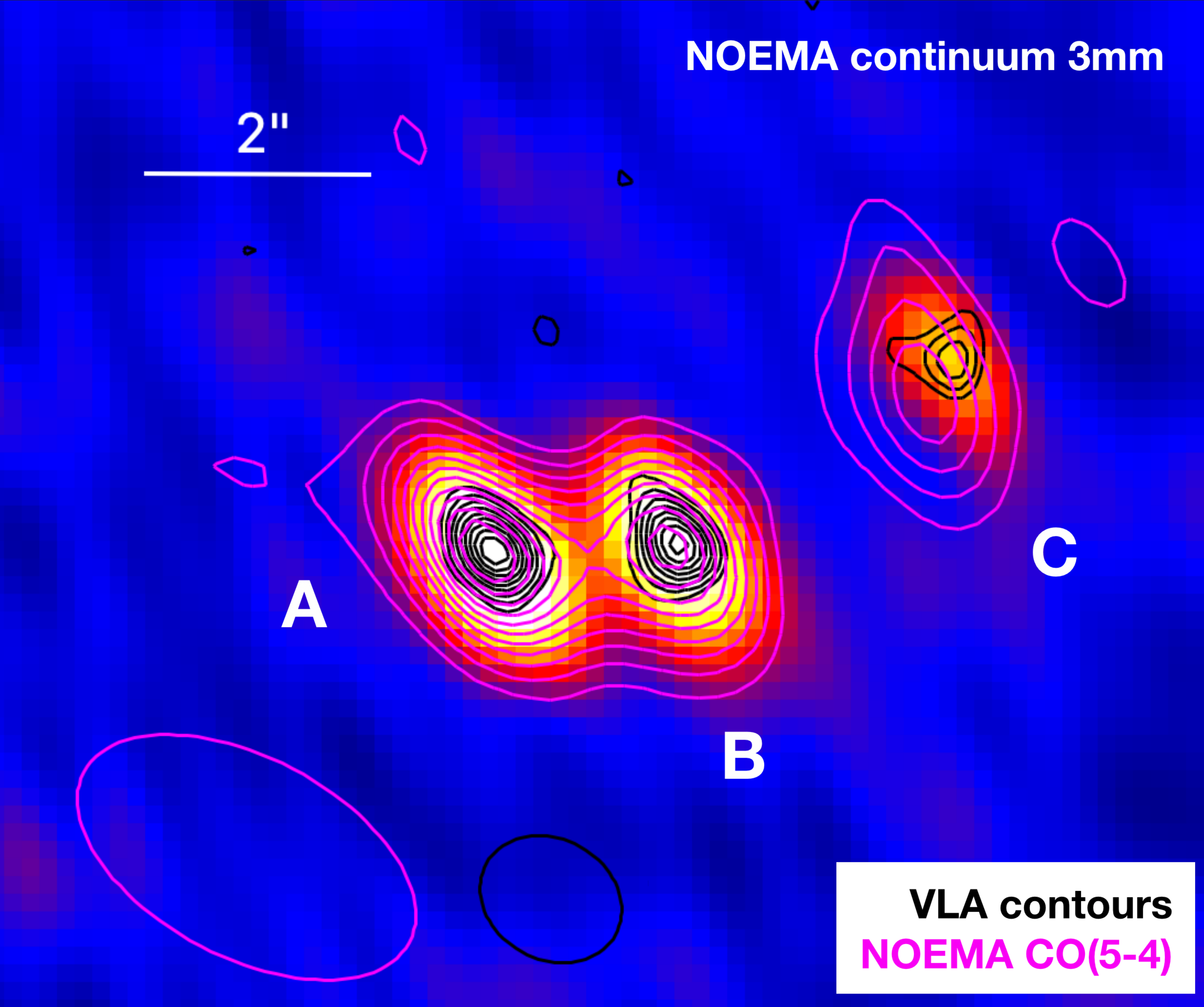}
  	\caption{ \label{noema_img} NOEMA 3\,mm continuum observation of RV4. The CO(5-4) emission contours are shown in magenta while VLA contours are shown in black for comparison. The VLA and NOEMA beams are indicated by the black and magenta ellipses, respectively.
	}
\end{figure}

RV4 was originally detected in the field of view of the \textit{Herschel}/SPIRE pointing observation of IC\,3521 (Fig.~\ref{imagessub}, left panel).
RV4 is unresolved in the three bands, and the SPIRE flux densities are 78.4$\pm$8.9, 118.4$\pm$10.9, and 122.8$\pm$11.1\,mJy at 250, 350, and 500\,$\mu$m, respectively.
The NIKA millimeter flux density ratio excludes the possibility of a low-$z$ radio AGN and confirms the high-$z$ nature of this emission with fluxes 21.1$\pm$1.2 and 6.0$\pm$0.2\,mJy at 1\,mm and 2\,mm, respectively, placing the IR peak of the SED between 350 and 500\microns.
The redshift of RV4 is provided by the detection of the CO(5-4), CO(4-3), and [CI]($^3P_1\rightarrow ^3P_0$) lines with 30\,m/EMIR (Fig.~\ref{emir}), yielding a spectroscopic redshift of 4.72401$\pm$0.00042.

Combining the SPIRE and NIKA observations and the spectroscopic redshift of RV4, we derive a total IR luminosity of 1.07$\pm$0.19$\times$10$^{14}$\,L$_{\odot}$ from IR SED fitting (see Sect.~\ref{irprop}).
The high S$_{500}$ flux density is above the 100\,mJy \cite{Negrello10} lens selection threshold, above which the probability for a SMG to be lensed is very high.
This strongly suggests the presence of multiple sources contributing to the submm fluxes \citep[e.g.,][]{Hodge13} or a magnification from lensing \citep[e.g.,][]{Negrello10}.
The SPIRE/NIKA emission is resolved by VLA A-configuration observations at 3\,GHz/10\,cm into three sources, hereafter named A, B, and C (Fig.~\ref{imagessub}, right panel).
The measured 10\,cm continuum flux densities are 34.3$\pm$4.6, 31.2$\pm$4.9, and 19.8$\pm$6.3\,$\mu$Jy for sources A, B, and C, respectively.

We determine the spectroscopic redshift of each of the three blobs from NOEMA/PolyFix observations using the CO(5-4) emission line.
The [CI](1-0) line is also detected but only for blob A and B; the [CI] to CO(5-4) line ratio is equivalent for both blobs, at 0.27$\pm$0.04 and 0.26$\pm$0.04 for blob A and B, respectively.
For each blob, and for the CO(5-4) and [CI](1-0) lines, we measure the redshift using the \textsc{slinefit}\footnote{\url{https://github.com/cschreib/slinefit}} code \citep{Schreiber18}. 
As shown in Fig.~\ref{noemalines}, the CO(5-4) lines of blob A and B are at nearly the exact same frequency (100.683\,GHz and 100.680\,GHz) with the same width (627$\pm$55\,km/s and 622$\pm$67\,km/s).
The redshift of blob C is slightly offset with a CO(5-4) line at 100.66\,GHz.
These frequencies translate into $z=4.72359^{+0.00022}_{-0.00011}$, $z=4.72375^{+0.00013}_{-0.00023}$, and $z=4.72469^{+0.00028}_{-0.00028}$, for blobs A, B, and C, respectively.
The errors quoted in the redshift measurements are underestimated, as the uncertainties from the line profiles are not included. 
Indeed, as discussed in different studies based on simulations \citep[e.g.,][]{Hezaveh12,Serjeant12} and observations \citep[e.g.,][]{Riechers08,Dye15,Yang17,YangC19,Apostolovski19}, the magnification can significantly vary along the velocity channels from blue to red and thus heavily distort the intrinsic line profile. 
This differential lensing effect may also cause different line profiles in different images.
This effect could be related to the redshift offset of blob C, which might be caused by a higher magnification of the blue part of the line. 
Although the [CI](1-0) line is weaker, the line frequencies of blobs A and B are consistent with a single redshift.
As for CO(5-4), the [CI](1-0) redshift of blob C seems to be offset, but in this case the line SNR is too low ($\approx$2) for a meaningful comparison.

In Fig.~\ref{noema_img}, we show that the positions of the NOEMA blobs are consistent with the VLA continuum positions.
The CO(5-4) positions of blobs A and B are also consistent with the continuum detections.
However, there seems to be a shift between the position of the CO(5-4) emission from blob C and its continuum counterparts.
The separation between the two centers of emission is 0.29\arcsec$\pm$0.10\arcsec.
In theory\footnote{see \url{https://www.iram.fr/IRAMFR/IS/IS2002/html_2/node131.html}}, for the CO(5-4) line, a 4.5 sigma detection has a position uncertainty of 0.17$\times$0.09\arcsec. 
The same uncertainty is expected for the continuum.
Therefore, this spatial shift might not be significant.

Considering the extreme IR luminosity of the system, the spatial distribution of the three blobs,and the identical CO(5-4) line profile (redshift and width) for each blob, we conclude that RV4 is a lensed $z=4.72$ submm galaxy, with A, B, and C, being multiple images of the same galaxy.

\begin{table*}
        \centering
        \tiny
        \caption{Submillimeter lines properties of RV4.}
        \begin{tabular}{l c c c c c c c c c}
        \hline\hline
        		 	& \multicolumn{3}{c}{Whole system} &blob A & blob B &blob C\\ 
        Line 		& $\nu_{obs}$	&	I$_{line}$	&	$\Delta v$ & 	I$_{line}$	&		I$_{line}$	&	I$_{line}$	 \\ 
        \hline
        			& GHz		&	Jy km/s&	km/s & 	Jy km/s&	Jy km/s& 	Jy km/s \\ 
        \hline
	    [CI](1-0)&	86.042		&	1.93$\pm$0.53	&	647$\pm$182 & 0.35$\pm$0.08&  0.23$\pm$0.09&  $<$0.36 (3$\sigma$) \\
	    CO(4-3)	& 	80.602	&	3.13$\pm$0.41 &	291$\pm$37 & - & -   &- \\
	    CO(5-4)	&	100.746		&	4.72$\pm$0.54 &	364$\pm$47 & 1.31$\pm$0.11&  0.89$\pm$0.1& 0.59$\pm$0.13 \\
	    \hline
        \label{sublines}
        \end{tabular}
\end{table*}

\begin{table*}
        \centering
        \tiny
        \caption{Submillimeter photometry of RV4.}
        \begin{tabular}{l c c c c c c }
        \hline\hline
        $\lambda$	& Instrument & Whole system & blob A  & blob B & blob C \\ 
        \hline
        	&  & mJy & mJy  & mJy & mJy \\ 
        \hline
        12\microns &   WISE 3      & $<$0.52 (3$\sigma$)&	-  & - & -\\
        22\microns &   WISE 4      & $<$3.24 (3$\sigma$)&	-  & - & -\\
        100\microns &   \textit{Herschel}/PACS      & $<$107 (3$\sigma$)&	-  & - & -\\
        100\microns &   \textit{Herschel}/PACS      & $<$188 (3$\sigma$)&	-  & - & -\\
	    250\microns	&	\textit{Herschel}/SPIRE		&	78.4$\pm$8.9    &	-  & - & -\\
	    350\microns	&	\textit{Herschel}/SPIRE		&	118.4$\pm$10.9  &	-  & - & -\\
	    500\microns	&	\textit{Herschel}/SPIRE		&	122.8$\pm$11.1  &	-  & - & -\\
	    1.2\,mm           &	NIKA		                & 21.1$\pm$1.2  & 	-  & - & -\\
	    2\,mm           &	NIKA		                &	6.0$\pm$0.2 &	-  & - & -\\
	    3\,mm           &	NOEMA		                &	-           & 0.416$\pm$0.020	  & 0.376$\pm$0.018  & 0.248$\pm$0.017\\	    
	    10\,cm          &	VLA		                    &	0.0783$\pm$0.0102 &	0.0343$\pm$0.0046  & 0.0312$\pm$0.0049 & 0.0198$\pm$0.0063\\
	    \hline
        \label{phot}
        \end{tabular}
\end{table*}

\subsection{Identification of the lens}
\begin{figure}%[!h] 
 	\includegraphics[width=\columnwidth]{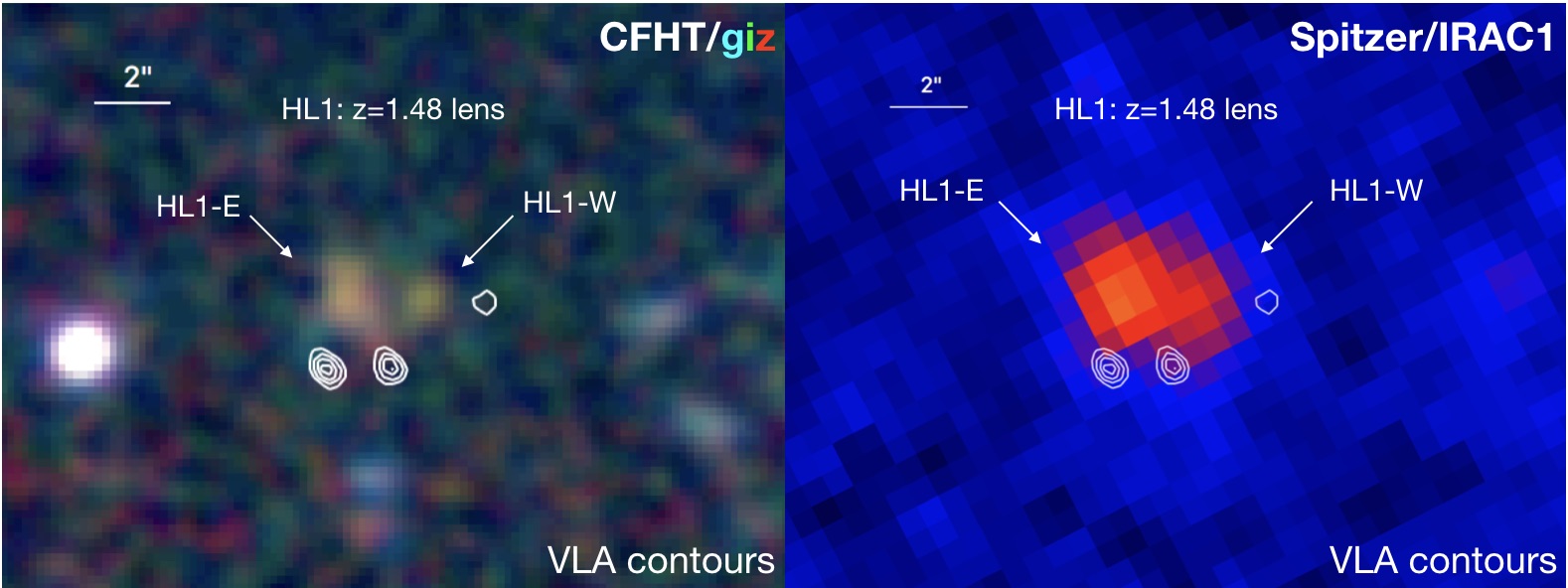}
    \caption{ \label{images} \textbf{Left panel:} CFHT/$g'$, $i'$, and $z'$ band images shown as a RGB image at the position of RV4. \textbf{Right panel:} \textit{Spitzer}/IRAC1 image centered on the position of RV4. In both panels, the white contours indicate the positions of the three VLA detections. The positions of the $z=1.48$ lens system are indicated by the white arrows.
    }
\end{figure}

Two sources, hereafter denoted as High-z Lens 1 West (HL1-W) and High-z Lens 1 East (HL1-E), are clearly detected in the $u^*$, $g'$, $i'$, and $z'$ bands.
The two sources, which are only 2" apart, are slightly offset from the three VLA and IRAM/NOEMA blobs, by approximately 2.2" to the north (Fig.~\ref{images}, left panel).
HL1-W and HL1-E are also clearly detected in IRAC imaging at 3.6 and 4.5\microns\ (Fig.~\ref{images}, right panel).
    
The $u^*$ and $g'$ bands probe shorter wavelengths than the Lyman break at $z=4.72$; thus we do not expect any emission from RV4 in these bands.
Moreover, as shown by the VLA contours on the same figure, no $i'$ or $z'$ emission is seen from these three blobs either.
The extended source detection limits of the NGVS are 26.3 and 25.8\,AB\,mag\,arcsec$^{-2}$ (2$\sigma$) for the $i'$ and $z'$ band, respectively \citep{Ferrarese12}.
In the \textit{Spitzer}/IRAC images, although the VLA detections are close to the outskirts of the IRAC emission of the lens system, no strong emission from RV4 is detected (Fig.~\ref{images}, right panel).
Furthermore, the IRAC fluxes measured from PSF fitting for HL1-W and HL1-E are consistent with its SED, indicating no particular excess of flux that could be attributed to RV4.

\subsection{Redshift determination for the lens}
    
\begin{figure*}%[!h] 
  	\includegraphics[width=\textwidth]{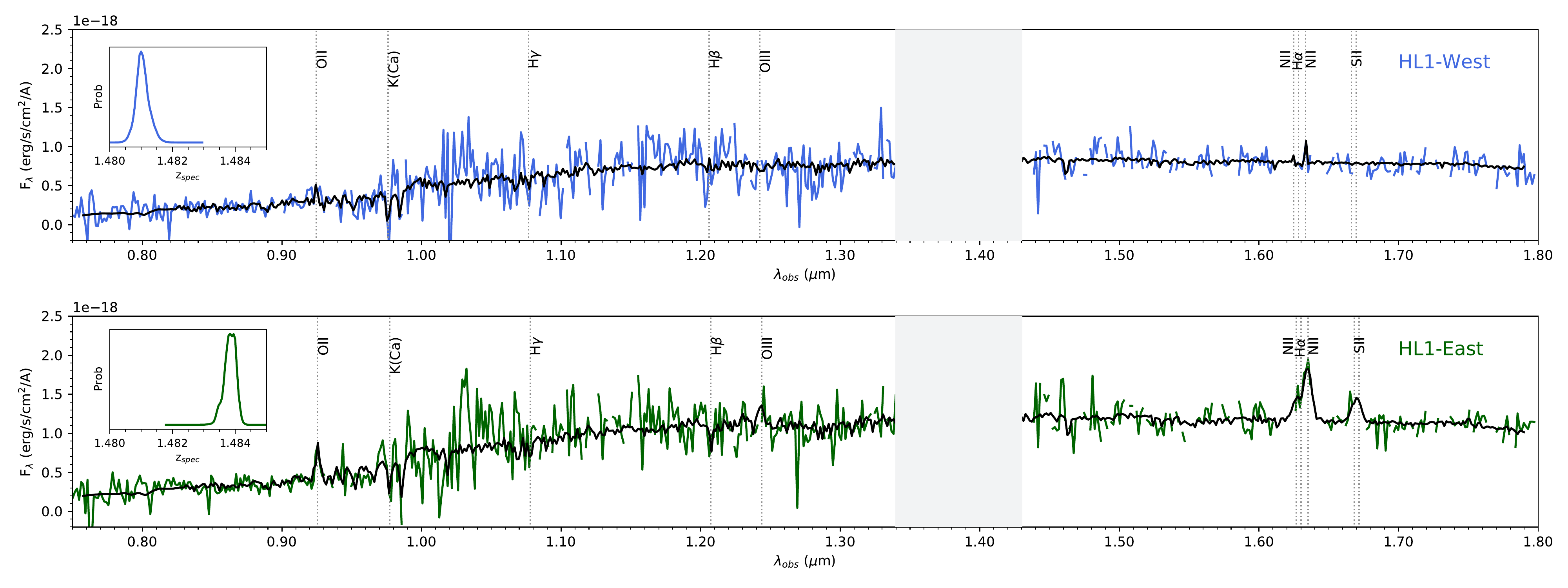}
  	\caption{ \label{xshooter} \textbf{Upper panel:} VLT/XSHOOTER 1D spectra of HL1-West. \textbf{Lower panel:} VLT/XSHOOTER 1D spectra of HL1-East. The inset panels show the probability distribution function for the redshift. The black lines are the best spectral fits obtained with \textsc{slinefit}.
	 }
\end{figure*}

To determine the redshift of HL1-W and HL1-E, we obtained VLT/XSHOOTER observations of the system (Fig.~\ref{xshooter}).
The spectroscopic redshifts are measured using the software \textsc{slinefit}.
For HL1-E and HL1-W respectively, we obtain $z_{spec}=$1.48379$^{+0.00018}_{-0.00019}$ and $z_{spec}=$1.48095$^{+0.00025}_{-0.00015}$, where the errors corresponds to uncertainties on the fits.
The small errors on these spectroscopic redshifts are driven by high signal-to-noise (SNR) detections of emission lines -- the [OII] doublet (SNR=7.7), [OIII] doublet (SNR=3.3), [NII] doublet (SNR=11.4), H$\alpha$ (SNR=5.3), and [SII] doublet (SNR=4.7) -- as well as a detection of the continuum emission.
These values reveal HL1-E/W as one of the most distant lens known in strong lensing systems and correspond to a velocity difference of 343$\pm$38\,km/s indicating that HL1-E and HL1-W are probably merging together.
The projected distance between the two blobs of 2\arcsec\ (17.4\,kpc) is also consistent with merger scenario.
In the literature, we find only one lens at a higher redshift presented in \cite{Canameras17} at $z_{spec}=1.525$, which is gravitationally lensing an SMG detected by \textit{Planck}.

\subsection{\label{sectlensprop}Lensing properties}
The data in hand allow us to obtain a better picture of the system in which RV4, at $z_{spec}=4.72$, is strongly lensed by a close pair of $z_{spec}=1.48$ galaxies.
The redshifts of these sources as well as the positions of the multiple images of RV4 (A, B, and C) are used as constraints to provide a first order lens model.
The redshifts of the lenses and sources as well as the centroids of images A, B and C are used to constrain a mass distribution assuming they arise from a multiply-imaged system. 
We use the \textsc{lenstool}\footnote{Publicly available at \url{https://projets.lam.fr/projects/lenstool/wiki/}} software  \citep{Kneib96,Jullo07} to optimize a parametric model of the mass distribution reproducing the locations of the lensed images. 
The model parameters are presented in Table~\ref{lensingtable}. 
The modeling have been made following a Occam's razor principle in the fact that this the simplest model able to correctly reproduce the positions of the VLA/NOEMA blobs as well as their flux ratios.
A thorough model comparison will be presented in the second paper.
Although the lens is composed of a pair of galaxies, we find that a simple model with a single dPIE (double Pseudo-Isothermal Elliptical) mass distribution provides a very good match, with an rms of 0.01\arcsec\ in the image plane. 
This simple model estimates the magnification to be  8.2$\pm$2.5 with an Einstein radius, constrained by the positions of blobs A, B, and C, of 2.2\arcsec$\pm$0.2.
The model provides magnifications of 3.4$\pm$1.9, 2.6$\pm$1.4, and 2.3$\pm$0.8 for blobs A, B, and C, respectively.

\begin{figure}%[!h] 
  	\includegraphics[width=\columnwidth]{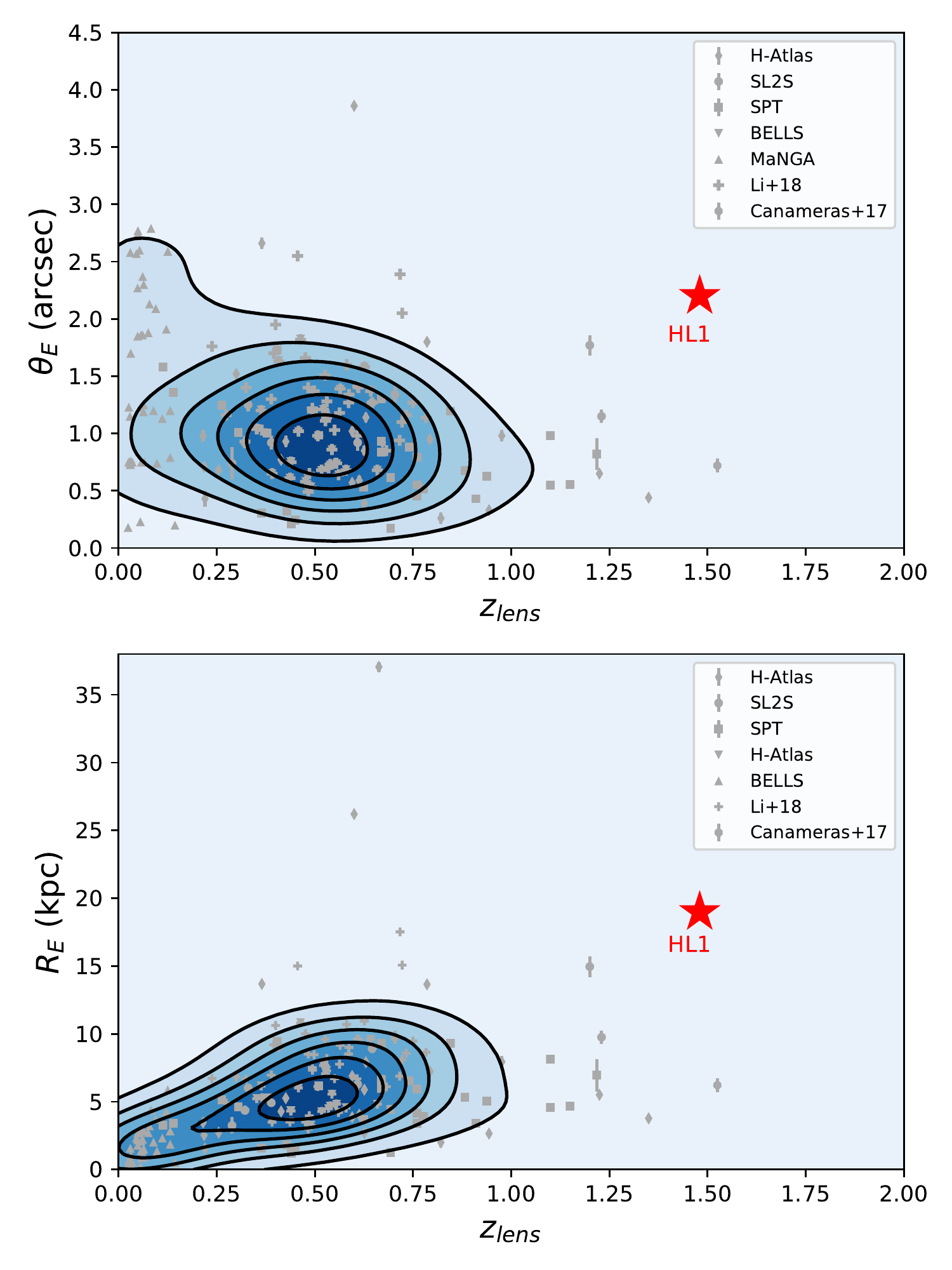}
  	\caption{ \label{lensprop}Einstein radii of lenses from the literature as a function of their redshift. The samples from the literature are: SL2S \citep{More12}, SPT \citep{Spilker16}, BELLS \citep{Shu16}, MaNGA \citep{Talbot18}, H-ATLAS \citep{Amvrosiadis18}, and the sources of \cite{Li18} and \cite{Canameras17} . The position of HL1 is marked by the red star. \textbf{Upper panel:} Einstein radius in arcsecond as a function of the redshift of the lens. Different symbols are from different samples of the literature. \textbf{Lower panel:} Same as above, but with the Einstein radius expressed in kiloparsec. The density contours are built from the same sample of literature sources.
	}
\end{figure}

We compare the redshifts and Einstein radii of other strong lensing systems found in the literature in Fig.~\ref{lensprop}.
In addition to being one of the most distant lenses known, HL1 has an Einstein radius of 2.2\arcsec$\pm$0.2 (20\,kpc at $z=$1.48) that places it in a new region of the Einstein radius versus redshift plane (Fig.~\ref{lensprop}).

\begin{table*}
        \centering
        %\tiny
        \caption{Lensing modeling parameters.}
        \begin{tabular}{l c c c c c c c}
        \hline\hline
        Model & $\Delta\alpha^{a}$ & $\Delta\delta^{a}$ & $e^{b}$ & $\theta^{c}$ & r$_{\rm core}^{d}$ & r$_{\rm cut}^{e}$ & $\sigma^{f}$ \\
        & [arcsec] & [arcsec] & & [deg] & kpc & kpc & km/s \\
        \hline 
        DM1 & $  0.2^{+  0.1}_{ -0.1}$ & $  0.4^{+  0.1}_{ -0.4}$ & $ 0.80^{+ 0.02}_{-0.25}$ & $-69.8^{+  0.9}_{ -0.9}$ & $8^{+4}_{-1}$ & $[500]$ & $436^{+34}_{-10}$ \\
        \hline
        \label{lensingtable}
        \end{tabular}
	    \tablefoot{
	    \tablefoottext{a}{Center position from an arbitrary reference taken at RA:188.68699 and Dec:+7.1356514.}
	    \tablefoottext{b}{Potential ellipticity, defined as $(a^2-b^2)/(a^2+b^2)$.}
	    \tablefoottext{c}{Potential angle.}
	    \tablefoottext{d}{Core radius of the dPIE potential.}
	    \tablefoottext{e}{Cut radius, fixed to 500\,kpc as it is not constrained from our data.}
	    \tablefoottext{f}{Velocity dispersion.}
		}
\end{table*}

\subsection{\label{phyprop}Physical properties of the lens}   
    
To derive the physical properties of each component of HL1, we combine the photometric data (CFHT, FOURSTAR, \textit{Spitzer}) with the XSHOOTER spectra.
The XSHOOTER spectra are scaled using the $i'$ band flux density for each blob and then integrated into a set of 10 artificial filters.
We use the SED modeling code CIGALE \citep{Boquien19} to perform the SED fitting.
We assume a \cite{Salpeter55} IMF, a flexible delayed SFH \citep[as described in][]{Ciesla17}, \cite{BruzualCharlot03} models, and a \cite{Calzetti00} attenuation law.
We find that HL1-W and HL1-E have stellar masses of 2.4$\pm$0.4$\times$10$^{11}$ and 3.3$\pm$0.6$\times$10$^{11}$\,M$_{\odot}$, respectively.
The SED fitting performed on the whole system using the integrated photometry of the two components yields a total stellar mass of 6.2$\pm$1.0$\times 10^{11}$\,M$_{\odot}$. 

\begin{figure*}%[!h] 
  	\includegraphics[width=\textwidth]{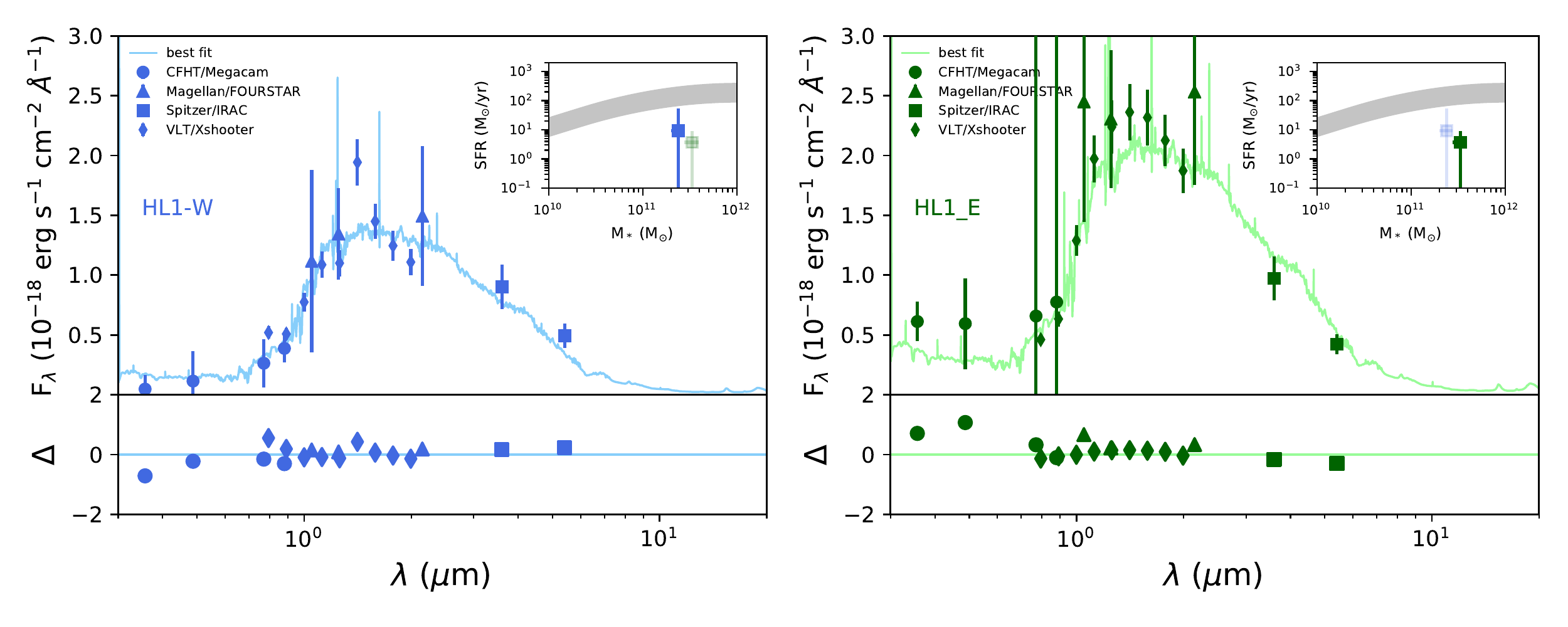}
  	\caption{ \label{sed_opt} SED modeling of the two components of HL1 (HL1-W: left panel, HL1-E: right panel). The fit is performed using the filled symbols, circles are photometric points, and diamonds are points obtained averaging the XSHOOTER spectra over a range of wavelengths. The best fit to the data is shown as a solid line. The subpanels show the difference between the data points and the best-fit models. In each inset panel, we show the position of the component (plain color) in the SFR-M$_*$ diagram along with the MS at $z=$1.5 from \cite{Schreiber15}. The position of the second component is also indicated in light color for comparison. Both components are located below the MS.
	  }
\end{figure*}

\begin{figure}%[!h] 
  	\includegraphics[width=\columnwidth]{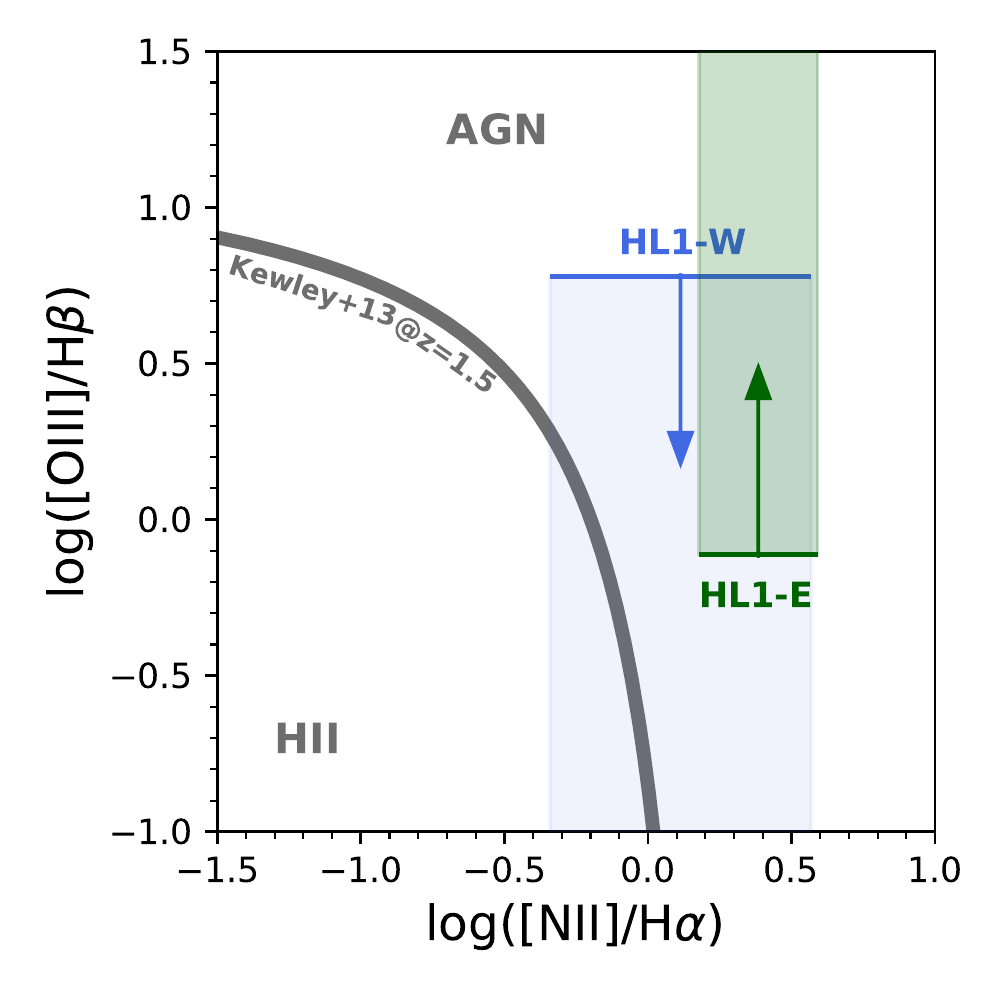}
  	\caption{ \label{bpt} HL1-W (blue) and HL1-E (green) placed on the BPT diagram. The limit of \cite{Kewley13} at $z=1.5$ is shown in grey.
	}
\end{figure}

Although the stellar mass is well constrained thanks to FOURSTAR and \textit{Spitzer} data, the star formation rate (SFR) estimates for HL1-W and HL1-E are more uncertain due to the faintness of the two galaxies.
The SED fitting results in SFRs of 9.2$^{+46.0}_{-9.2}$ and 3.7$^{+5.1}_{-3.7}$\,M$_{\odot}$\,yr$^{-1}$ for HL1-W and HL1-E, respectively, where the errors reflect the difficulty of estimating the SFRs from the faint $g'$ and $i'$ emission.
These values place the two counterparts below the MS of \cite{Schreiber15} at $z=1.5$, as shown in Fig.~\ref{sed_opt}.
However, we can probe the SF activity of these two sources using the H$\alpha$ emission detected in the XSHOOTER spectra.
As shown in Fig.~\ref{xshooter}, HL1-W has no H$\alpha$ emission, confirming the fact that HL1-W seems to be passive.
On the other hand, HL1-E seems to have a large and strong H$\alpha$ line with a line flux of 1.79$\pm$0.33$\times$10$^{-17}$\,erg\,s$^{-1}$\,cm$^{-2}$ associated with a strong [NII] line with a flux of 4.33$\pm$0.38$\times$10$^{-17}$\,erg\,s$^{-1}$\,cm$^{-2}$.
Furthermore, the H$\beta$ emission line is not detected whereas a strong [OIII] line with a flux of 2.40$\pm$0.74$\times$10$^{-17}$\,erg\,s$^{-1}$\,cm$^{-2}$ is detected.
These line fluxes, as well as the H$\beta$ upper limit, place HL1-E in the AGN region of the BPT diagram, as shown in Fig.~\ref{bpt}.
Since HL1-E is likely hosting an AGN, we check if our measured stellar mass can be contaminated by AGN emission in the NIR. 
To do this, we use CIGALE to quantify the AGN fraction, defined as the contribution of the AGN to the total L$_{IR}$ of the galaxy, which is a scaling factor for the AGN emission model.
The methodology is fully described in \cite{Ciesla15}.
For HL1-E, the best $\chi^2$ model yields $frac_{AGN}=0$, meaning that an AGN emission component is not needed to reproduce the host galaxy UV-NIR emission.
Therefore our stellar mass measurement is not contaminated by the AGN emission.
We conclude that HL1-W and HL1-E seem to be passive, as no strong star formation activity is detected from either continuum or line analysis.

For the SED modelling, we adopt a \cite{Calzetti00} attenuation law, which is better suited for star-forming galaxies than for passive galaxies.
However, \cite{Calzetti00} assume almost no attenuation in the NIR.
Other attenuation laws such as \cite{CharlotFall00} and \cite{LoFaro17} are characterized by a stronger attenuation in the NIR yielding larger stellar masses, by a factor of $\sim$1.3 \citep{LoFaro17}. 
 
%=================================================================================
\section{\label{rv4}RV4 as a starburst galaxy}

    \subsection{\label{irprop}Infrared properties}

    \begin{table*}
        \centering
        \caption{Physical properties of RV4.}
        \begin{tabular}{l l c c}
         \hline\hline
        Property & unit & \multicolumn{2}{c}{Value} \\ 
        \hline\hline
                    &   & \cite{Schreiber18} & \cite{Draine14} \\ 
	    $\mu$L$_{IR}$	& 10$^{14}$L$_{\odot}$	& 1.06 $\pm$0.06 &	1.07 $\pm$0.19\\
	    T$_{dust}$	& K			& 41.3 $\pm$1.2	&	35.7$\pm$11.6\footnotemark[1]	\\
	    $\mu$M$_{dust}$	& 10$^{9}$M$_{\odot}$	&  &	10.7 $\pm$0.4\\
	    q$_{IR}$		& -			& 2.70 $\pm$ 0.09	& 2.69 $\pm$ 0.09		\\
	    $\alpha$		& -			& 0.70 $\pm$ 0.08	& 0.68 $\pm$ 2.08	\\
	    \hline
	    $\mu$M$^{CI}_{gas}$ &10$^{11}$ M$_{\odot}$ &  \multicolumn{2}{c}{2.7 $\pm$ 0.7}\\
	    $\mu$M$^{CO}_{gas}$ &10$^{11}$ M$_{\odot}$ &  \multicolumn{2}{c}{3.1 $\pm$ 0.4}\\
	    $\mu$M$^{dust}_{gas}$ & 10$^{11}$ M$_{\odot}$ &  \multicolumn{2}{c}{6.0 $\pm$ 3.0} \\
	    t$^{CI}_{depl}$ & Myr & \multicolumn{2}{c}{14.4$\pm$4.7}\\
	    t$^{CO}_{depl}$ & Myr & \multicolumn{2}{c}{17.2$\pm$2.2}\\
	    t$^{dust}_{depl}$ & Myr & \multicolumn{2}{c}{32.9$\pm$18.7}\\
	    \hline
	    $\mu$L'$_{CI}$ & 10$^{10}$ K km s$^{-1}$ pc$^2$& \multicolumn{2}{c}{9.6$\pm$2.6}\\
	    $\mu$L'$_{CO(4-3)}$ & 10$^{10}$ K km s$^{-1}$ pc$^2$ & \multicolumn{2}{c}{17.8$\pm$2.3}\\
	    $\mu$L'$_{CO(5-4)}$ & 10$^{10}$ K km s$^{-1}$ pc$^2$ & \multicolumn{2}{c}{17.1$\pm$1.9}\\
	    \hline
	    $\Sigma_{IR}$ & 10$^{11}$ L$_{\odot}$kpc$^{-2}$ & \multicolumn{2}{c}{7.7$\pm$1.6} \\
	    \hline\hline
        \label{irproptable}
        \end{tabular}
        \footnotetext[1]{The temperature was obtained from $U_{min}$ following \cite{Aniano12}.}
    \end{table*}

    \begin{figure}%[!h] 
  	    \includegraphics[width=\columnwidth]{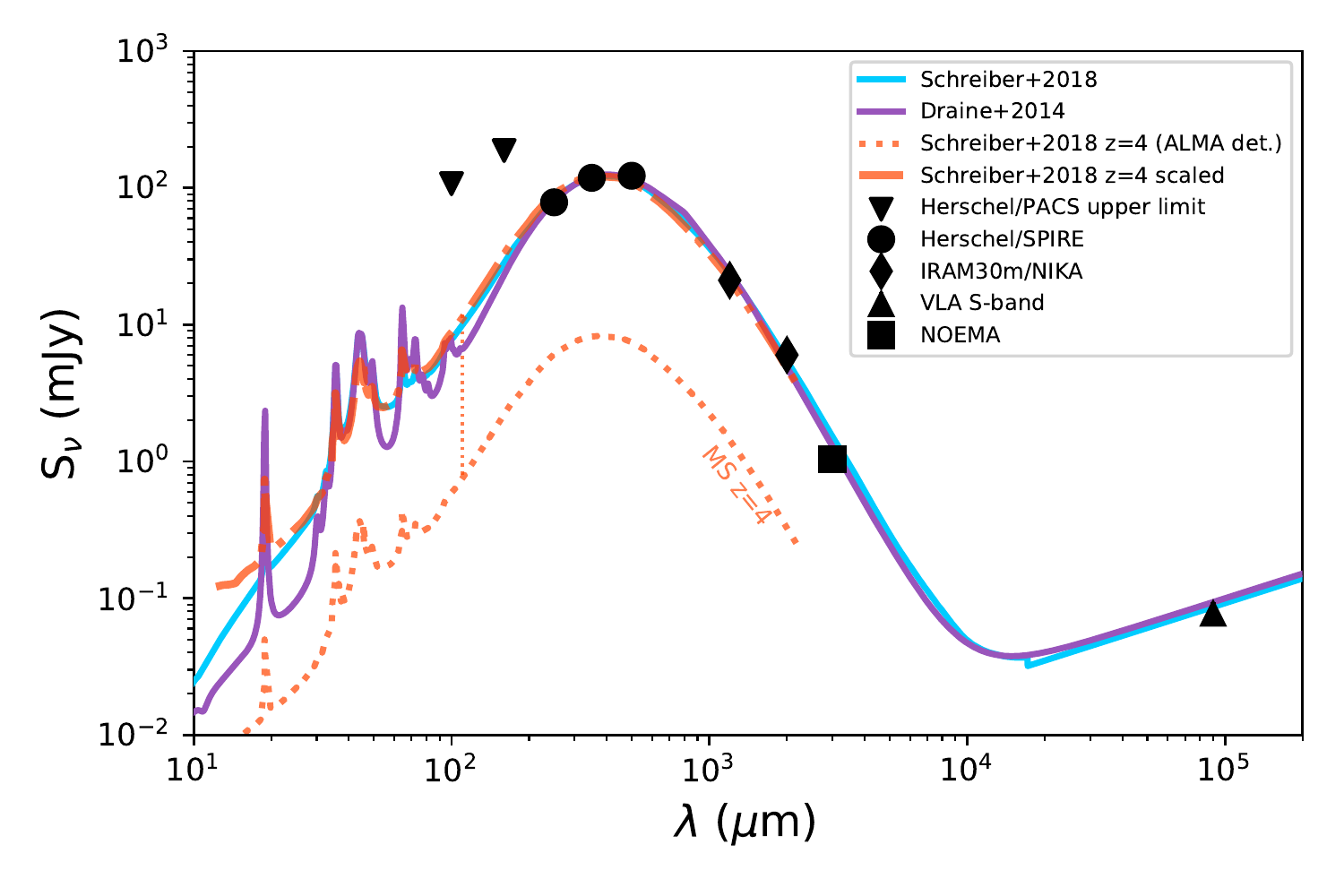}
  	    \caption{ \label{sed} IR SED modeling of RV4. The flux densities are not corrected from magnification. Black circles are the SPIRE data, while diamonds and triangle are the NIKA and VLA fluxes, respectively. The downward triangles indicate the non detection from PACS. The IR SED was fit using the IR libraries of \cite{Schreiber18} (light blue) and \cite{Draine14} (purple). The MS IR SED derived by \cite{Schreiber17} from a sample of ALMA detected $z=4$ sources is shown with the dashed orange line. A scaling of this MS SED to the 350\microns\ flux of RV4 leads to a scaling factor of 14.8.
	}
    \end{figure}
    
    The combination of the three SPIRE detections with the NIKA 1\,mm and 2\,mm flux densities provides a good sampling of the IR SED of RV4.
    Using the SED modeling code CIGALE, we fit the IR SED using two different dust emission models.
    The first is the library of \cite{Draine14}, which is an extension of the \cite{DraineLi07} dust emission model, while
    the second set of models is presented in \cite{Schreiber18}. 
    As shown in Fig.~\ref{sed}, both libraries provide a good fit to the data, with $\chi^2_{red}$ of 2.94 and 3.77 for \cite{Schreiber18} and \cite{Draine14}, respectively.
    The $L_{IR}$ obtained with the two libraries is $\log(\mu L/L_{\odot})$ = 14.03 (Table~\ref{irproptable}). 
    As a free parameter of the fit, the temperature obtained from the \cite{Schreiber18} model is 41.3$\pm$1.2\,K. 
    The library of  \cite{Draine14} does not directly fit the temperature, but the parameter $U_{min}$ can provide an approximation of the temperature through the formula derived by \cite{Aniano12}, $T_{dust}\sim20U_{min}^{0.15}$\,K, leading to a temperature of 35.9$\pm$9.7\,K.
    At $z=4.72$, given the dust temperature measured, the Cosmological Microwave Background (CMB) should have a negligible impact on the L$_{IR}$ of RV4 \citep{DaCunha13}.
    From the SED fitting, we also estimate $\log(\mu M_{dust}/M_{\odot})=$10.03.

\begin{figure}%[!h] 
  	    \includegraphics[width=\columnwidth]{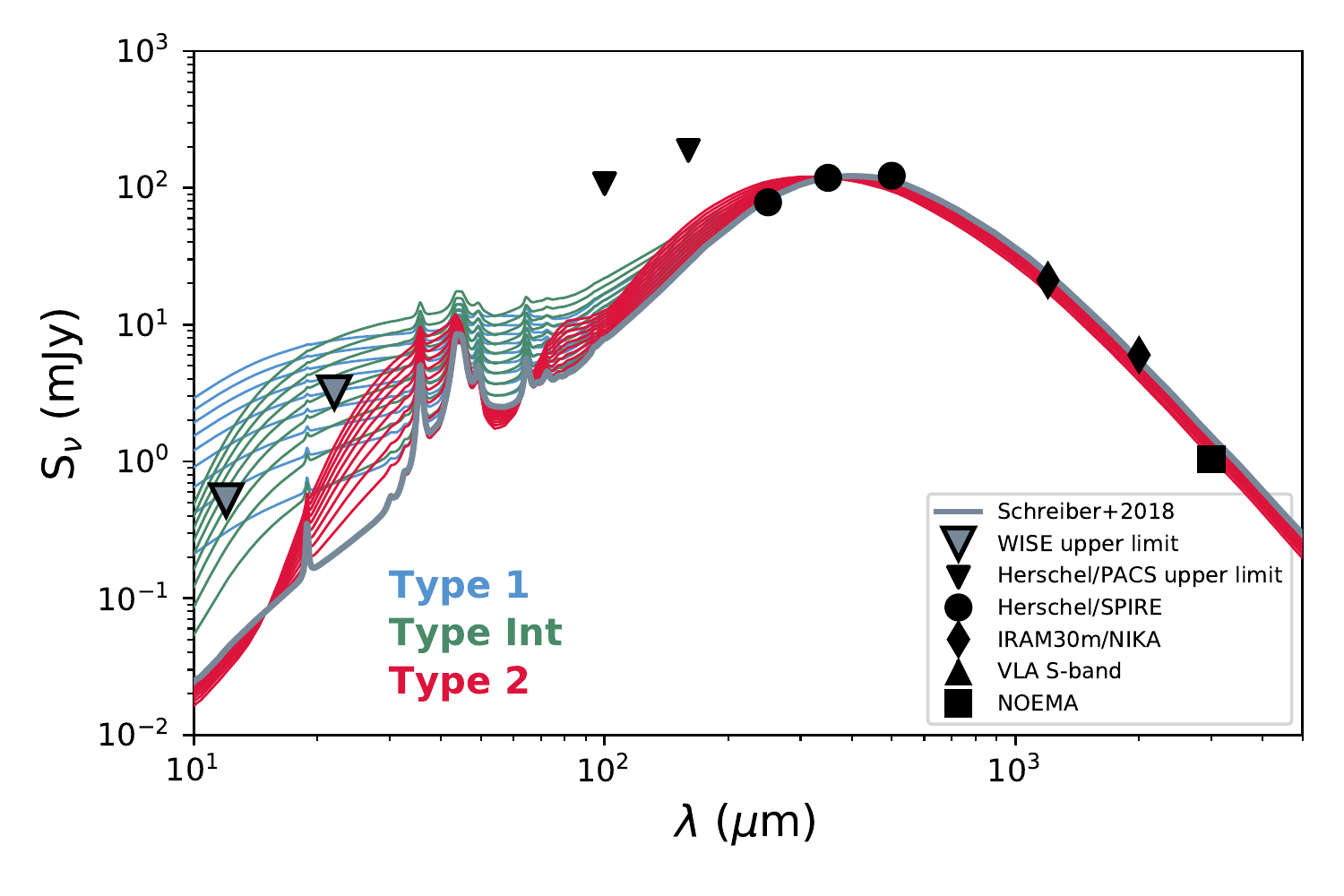}
  	    \caption{ \label{agn} Addition of different AGN contributions to the IR SED of RV4. AGN types simulated are type 1 (blue), intermediate type (green), and type 2 (red), as defined in \cite{Ciesla15}. The different lines correspond to different power of the AGN from 0\% (no AGN) to 50\% of the total IR luminosity. The gray downward triangles are the WISE 12 and 22\microns\ upper limits while the black downward triangles are the \textit{Herschel}/PACS upper limits.
	}
    \end{figure}

    It has been found that $\sim$20$\%$ of SMGs host an AGN \citep[e.g.,][]{Coppin10}; therefore, we investigate if our data allow us to rule out the presence of an AGN contributing to the IR luminosity of RV4.
    The SMG is not dectected in the PACS images of IC3521 at 100 and 160\microns. 
    The 3$\sigma$ upper limits determined as in \cite{Ciesla12} and \cite{Cortese14} do not place any useful constraints on the IR SED of RV4, as shown in Fig.~\ref{sed}.
    However, we have retrieved WISE 12 and 22\microns\ maps of the region around RV4. 
    The SMG is not detected, so we use the 3$\sigma$ detection limits provided by the explanatory supplement\footnote{\url{http://wise2.ipac.caltech.edu/docs/release/allsky/expsup/sec6_3a.html}}.
    Different models were assumed, and thus different values are quoted for the sensitivity limits.
    We assume RV4 to be unresolved in WISE bands, and take the less constrictive upper limit provided in the explanatory supplement, i.e., 0.52 and 3.24\,mJy (3$\sigma$) at 12 and 22\microns, respectively, as shown in Fig.~\ref{agn}.
    Using the best fit of the IR SED obtained in Sect.~\ref{irprop}, we add different AGN components assuming a large range of AGN fraction, defined here as the AGN contribution to the total IR luminosity \citep{Ciesla15}.
    As shown in Fig.~\ref{agn}, our data do not allow us to exclude the presence of a dust obscured, type 2 AGN that could contribute to the IR luminosity of RV4.
    However, the two WISE upper limits indicate that, were RV4 hosting a type 1 or intermediate type of AGN, we should expect a contribution to the total L$_{IR}$ less than 28$\%$.
    We note that at these wavelengths, considering the redshift of RV4, we may need to take the stellar emission into account, therefore, the 28$\%$ contribution should be considered as an upper limit.
     As a final test, we determine the IR/radio coefficient, q$_{IR}$, to understand if RV4 could be a radio-loud AGN.
    Using the VLA continuum data point (Fig.~\ref{sed}), we obtain a $q_{IR}=$2.70$\pm$0.09 using the CIGALE code.
    This is a typical value for star-forming galaxies \citep{Seymour09,Sargent10}.
    We therefore conclude that RV4 is not a radio-loud AGN, although the data in hand do not allow us to place a further constraint on the possible presence of a dust enshrouded AGN.
    
    \subsection{\label{gas}Gas mass and depletion time}
    Given that we detect high-J transitions of CO, i.e. CO (5$\rightarrow$4) and CO(4$\rightarrow$3), deriving a gas mass requires that uncertainties in the spectral line energy distribution (SLED) must be forded in with the uncertainty on the adopted $\alpha_{CO}$ value.
    Therefore, we first use the [CI] flux measurement to derive the H$_2$ mass.
    Indeed, \cite{Bothwell17} showed that [CI] and CO(2-1) have similar kinematic properties in a sample of dusty star-forming galaxies, suggesting that [CI] traces the same gas component as low-J CO emission.
    Furthermore [CI] as a gas mass tracer is far less affected by metallicity than CO \citep[e.g.,][]{Bothwell13}, although it does require an assumption that the line is optically thin.
    
    We follow the prescription of \cite{PapadopoulosGreve04}:
    \begin{equation}
    \frac{M_{H_2}^{CI}}{\mathrm{M_{\odot}}} = 4.92 \times 10^{10} h'^{-2} k(z) \left(\frac{X_{CI}}{10^{-5}} \right)^{-1}
    \left( \frac{A_{10}}{10^{-7} \mathrm{s^{-1}}}\right)^{-1} Q_{10}^{-1} \frac{S_{CI}}{\mathrm{Jy\,km\,s^{-1}}},
    \end{equation}
    with
    \begin{equation}
    k(z) = \frac{(1 + z - \sqrt{1+z})^2}{1 + z},
    \end{equation}
    \noindent and where $h'$=0.75, the Einstein A-coefficient $A_{10}$=7.93$\times$10$^{-8}$\,s$^{-1}$, $X_{CI}$=3$\times$10$^{-5}$, and $Q_{10}$=$Q_{10}(n,T_k)$ depending on $n$ and $T_k$ the gas density and temperature \citep{Weiss03,Papadopoulos04}.
    Assuming the abundance and excitation factors of \cite{Alaghband13}, we find $\mu$M$_{H_2}^{CI}$=1.85$\pm$0.51$\times$10$^{11}$\,M$_{\odot}$, with $\mu$ the magnification factor due to lensing, which yields $\mu$M$_{gas}^{CI}$=2.65$\pm$0.73$\times$10$^{11}$\,M$_{\odot}$, assuming a 36$\%$ helium contribution and no significant atomic H component as in \cite{Yang17}.
    Converting the $L_{IR}$ to SFR we obtain $\mu$SFR$=$1.82$\times$10$^{4}$\,M$_{\odot}$\,yr$^{-1}$ and derive a depletion time as:
    \begin{equation}
    t_{dep} = \frac{\mu M_{gas}^{CI}}{\mu \mathrm{SFR}},
    \end{equation}
    \noindent finding $t_{dep}$=14.4$\pm$4.7\,Myr.
    This is a short depletion time highly suggestive of a high-efficiency, bursty star formation, typical for SMGs \citep[e.g.,][]{Aravena16}.

    However, as the gas mass measurement obtained from CI depends on the assumption that CI is optically thin, to provide an independent check, we also derive the gas mass using the CO(4-3) line (Table~\ref{irproptable}). 
    Assuming $L'_{CO(4-3)}/L'_{CO(1-0)}=0.46$ \citep{CarilliWalter13}, we obtain $\mu L'_{CO(1-0)}=3.9\pm0.5\times$10$^{11}$\,K\,km\,s$^{-1}$\,pc$^2$.
    Using $\alpha_{CO}=0.8$\,M$_{\odot}/$K\,km\,s$^{-1}$\,pc$^2$, we obtain $\mu$M$_{gas}^{CO}=3.1\pm0.4\times$10$^{11}$\,M$_{\odot}$, which is consistent with the estimate obtained from CI.
    This gas mass estimate yields a depletion time of 17.2$\pm$2.2\,Myr.
    Although the uncertainty of this estimate does not reflect the uncertainty in the $L'_{CO(4-3)}/L'_{CO(1-0)}$ ratio and $\alpha_{CO}$ assumptions, it is consistent with the value obtained from CI.

    Finally, we use the dust mass as a third independent gas mass indicator:
    \begin{equation}
    \mu M_{gas}^{dust} = \delta_{GDR} \times \mu M_{dust},
    \end{equation}
    \noindent with $\delta_{GDR}$ the gas-to-dust mass ratio.
    We assume here the average $\delta_{GDR}$ of 56 ($\pm$28) obtained by \cite{Yang17} for a sample of lensed SMG.
    We obtain $\mu$M$_{gas}^{dust}$=6.0$\pm$3$\times$10$^{11}$\,M$_{\odot}$ and from this a $t_{dep}$ of 32.9$\pm$18.7\,Myr.
    Thus, using the dust mass, we obtain a larger, yet still small value of the depletion time indicating a starbusting phase.
    Using the $\delta_{GDR}$ obtained from the ALESS sample \citep{Simpson14,Swinbank14}, 75$\pm$10, would yield a depletion time of 44.0$\pm$5.9\,Myr.
    However, the sample of lensed SMGs of \cite{Yang17} has physical properties that are close to those of RV4, therefore we use the gas mass obtained using the gas-to-dust ratio of \cite{Yang17} in the following.
    These derived values are short, comparable to those of other starburst galaxies, and pointing towards a rapid starburst episode.
    We note that when computing the depletion times, we assume that the magnification $\mu$ is the same for the gas component and the IR produced by star-formation since we cannot make a detailed investigation of differential magnification with the data in hand.
    
    Given the good coverage of RV4's IR SED with 6 data points from 250\microns\ to 3\,mm observed (44 to 524\microns, rest frame), we derived the dust mass of RV4 from SED modeling. 
    We used an updated version of the model of \cite{DraineLi07} \citep{Draine14}, which is physically motivated and takes into account the different contributions from the dust heated in PDR and the dust heated by more evolved stars. 
    As an alternative, \cite{Scoville16} proposed a method useful when only one observations is available in the Rayleigh-Jeans part of the SED but that relies on assumptions such as the dust temperature that can lead to large uncertainties \citep{Berta16}. 
    Given our good coverage of both the peak of the IR emission and the Rayleigh-Jeans part of the SED, we prefer to rely on our measurement based on SED modeling.
    However, for comparison, we compute the gas mass of RV4 obtained by the \cite{Scoville16} method and obtain 7.7$\pm$2.0$\times$10$^{11}$\,M$_{\odot}$, uncorrected from magnification.
    This leads to a depletion time of 42.8$\pm$11.1\,Myr.
    
    \subsection{Gas density}
\begin{figure}%[!h] 
       	\includegraphics[width=\columnwidth]{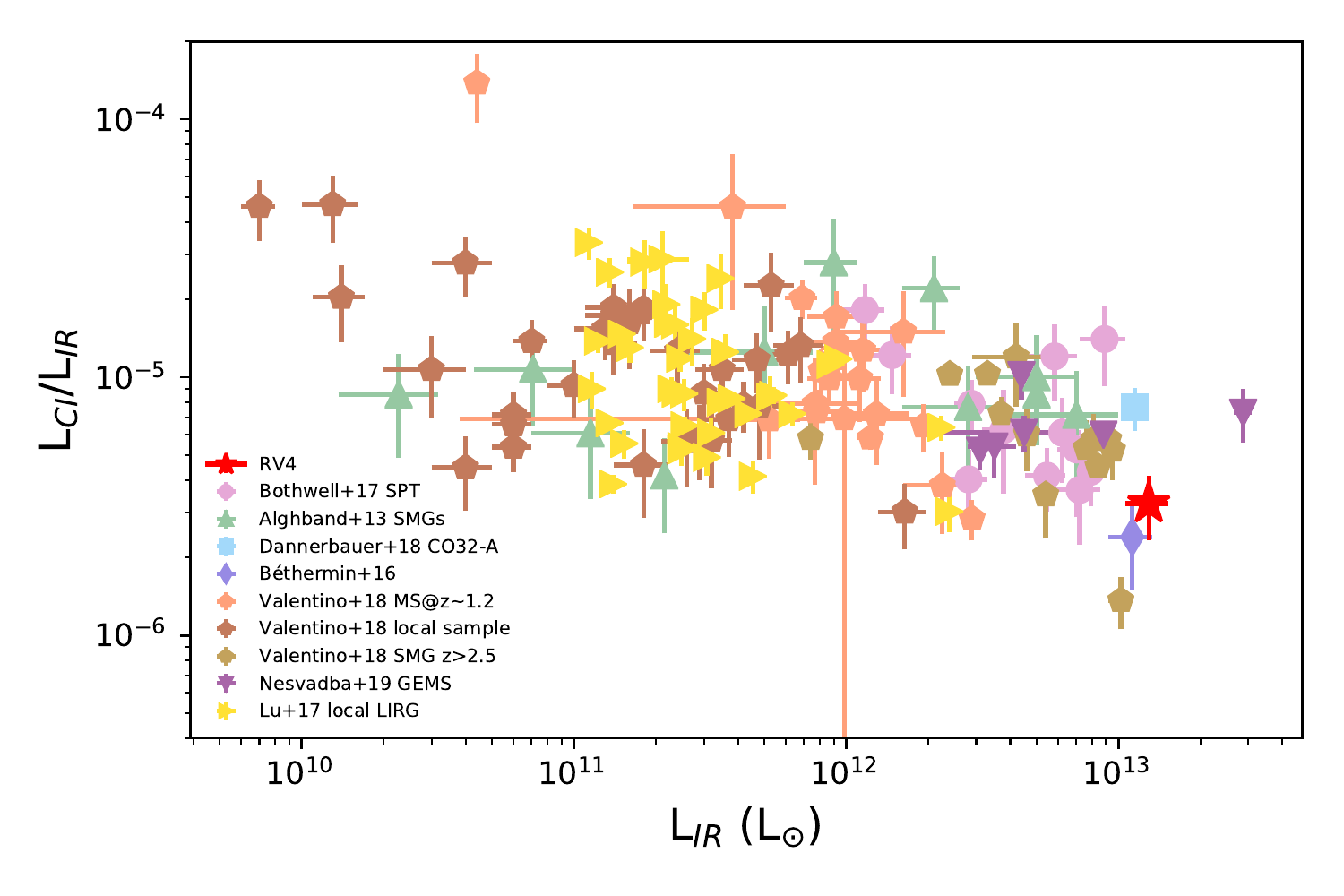}\\
  	    \includegraphics[width=\columnwidth]{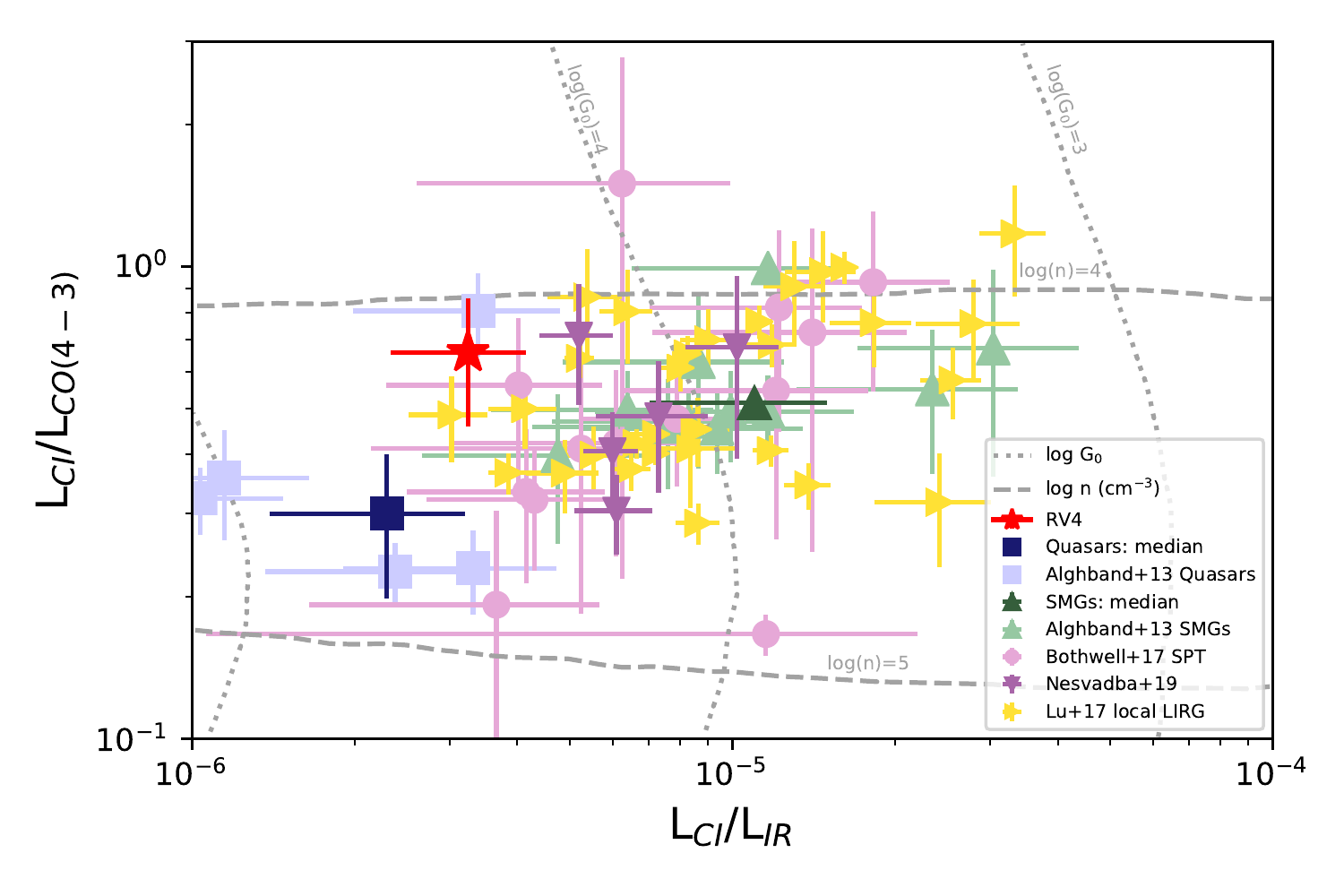}
  	    \caption{ \label{lci_lco} \textbf{Upper panel:} [CI] to LIR luminosity ratio (both in L$_{\odot}$) as a function of infrared luminosity, corrected for lensing magnification when needed, for several samples. The position of RV4 is indicated by the red star. For comparison we show the samples gathered by \cite{Valentino18} (high-z MS, SMG, and local galaxies), the SPT sample of \cite{Bothwell17}, the SMG sample of \cite{Alaghband13}, the galaxy CO32-A \citep{Dannerbauer18}, SPT-S J213242-5802.9 \citep{Bethermin16}, Planck's dusty GEMS \citep{Nesvadba19}, and the local LIRG sample of \cite{Lu17}. 
  	    \textbf{Lower panel:}  [CI] to CO(4-3) luminosity ratio versus [CI] to IR luminosity ratio (L$_{\odot}$). The position of RV4 is marked by the red star. This figure is adapted from \cite{Alaghband13}. Their SMG sample is shown (green triangles) as well as its median position (dark green triangle). The blue squares indicate the quasar sample presented in \cite{Walter11} with the dark blue square indicating the position of the median of their sample. The SPT sample of \cite{Bothwell17}, Planck's dusty GEMS \citep{Canameras18,Nesvadba19}, and the local sample of LIRGs studied in \cite{Lu17} are also shown in the figure. The grey lines indicate the contours of the gas density ($n$) and and the radiation field ($G_0$) for the corresponding $L_{CI}/L_{CO(4-3)}$ and $L_{CI}/L_{IR}$ ratios, as produced by the PDR models of \cite{Kaufman99}.  
	    }
    \end{figure}

    In this section, we use the computed IR, CO(4-3), and [CI](1-0) luminosities of RV4 obtained from EMIR (summing the emission of all three blobs) to investigate its gas properties.
    First, we present in Fig.~\ref{lci_lco} (top panel) the relation between [CI](1-0) to IR ratio and the IR luminosity.
    For comparison, we add a selection of data from the literature \citep{Walter11,Alaghband13,Bethermin16,Bothwell17,Lu17,Canameras18,Dannerbauer18,Valentino18,Nesvadba19}. 
    The position of RV4 is marked by the red star, taking into account the magnification from lens modeling.
    Even after this correction, RV4 is one of the most luminous IR sources of the compilation of SMGs shown in Fig.~\ref{lci_lco} (top panel).
    Its $L_{IR}$ is comparable to those of the recently published sources of \cite{Bethermin16} ($L_{IR}=1.1\pm0.2\times10^{13}\,L_{\odot}$) and of \cite{Dannerbauer18} ($L_{IR}=1.1\pm0.1\times10^{13}\,L_{\odot}$).
    Although the CO32-A source studied in \cite{Dannerbauer18} has a [CI](1-0) to IR luminosity ratio consistent with sources with a less extreme $L_{IR}$, RV4 and the source analyzed in \cite{Bethermin16} exhibit a lower ratio. 
    
    The locations of RV4, CO32-A, and the source of \cite{Bethermin16} in the upper panel of Fig.~\ref{lci_lco} seems to indicate a possible deviation from the main trend at high IR luminosity, especially if we also consider the $z>2.5$ SMG source of \cite{Valentino18} with the lowest $L_{CI}/L_{IR}$ ratio.
    However, more statistics is needed to confirm this.
    The [CI] luminosity traces the total gas while the IR luminosity traces the SF activity over a typical scale of 100\,Myr.
    Therefore, this ratio can be interpreted as a star formation efficiency indicator.
    The low $L_{CI}/L_{IR}$ ratio is thus an indication that star formation is highly efficient in RV4.
    
    To explore the properties of the ISM of submillimeter galaxies, \cite{Alaghband13} investigated the position of galaxies in the  $L_{CI}/L_{CO(4-3)}$ versus $L_{CI}/L_{IR}$ diagram.
    Although they acknowledged issues with simple PDR modeling, they compared a sample of quasars and SMGs to a set of PDR models from \cite{Kaufman09} and found that the SMGs have densities and radiation field strengths that are consistent with those of local starbursts, although $\sim$35\% of their SMG sample has lower density and weaker radiation field.
    They interpret this as an evidence that star formation can be extended in some submillimeter galaxies.
    The quasar sample is found at higher densities and radiation fields than SMGs on average.
    In Fig.~\ref{lci_lco} (bottom panel), we plot RV4 with the two samples of \cite{Alaghband13} on the $L_{CI}/L_{CO(4-3)}$ versus $L_{CI}/L_{IR}$ diagram, along with the SPT data of \cite{Bothwell17}.
    In terms of  $L_{CI}/L_{CO(4-3)}$ ratio, RV4 is consistent with what is found for SMGs, although the ratio is among the 25$\%$ highest for the two samples shown in Fig.~\ref{lci_lco} (lower panel).
    The position of RV4 in this diagram is close to the $log(n_H/cm^{-3})=4$ density line, indicating a rather low gas density compared to the other SMGs for the sample.
    However, in terms of  $L_{CI}/L_{IR}$, RV4 has the lowest ratio of the SMG sample, close to the values obtained for quasars.
    This is an indication of a rather intense radiation field, which is consistent with RV4 being a starburst.
    This low gas density is also confirmed by the $L'_{CO(5-4)}/L_{IR}$ ratio for RV4, which corresponds to only 54$\pm$11\% of the expected value for normal galaxies \citep{Daddi15}.

    \subsection{Size and IR surface brightness}
    
    In the VLA observations, with a beam size of 0.68\arcsec$\times$ 0.55\arcsec and a position angle of 30 degrees, all three blobs are compatible with being point sources.
    The emission from blob A and B can be deconvolved from the beam but with large uncertainty.
    Given the sensitivity (2.6\,$\mu$Jy/beam) of the VLA observations, measuring an accurate size is challenging, and we cannot exclude that the sources are marginally extended.
    However, we can measure the sizes of the three RV4 blobs in the $uv$ plane of the NOEMA data, combining all continuum channels.
    Assuming a circular Gaussian profile, the resulting observed sizes are 0.59\arcsec$\pm$0.05\arcsec, 0.50\arcsec$\pm$0.05\arcsec, and 0.50\arcsec$\pm$0.08\arcsec\ for blobs A, B, and C, respectively.
    All sources are clearly resolved and, within the uncertainties, have approximately the same size.
    We note that these sizes are not corrected for the shear that the strong lensing is causing.
    We also tried an elliptical Gaussian profile fit, but only the brightest source, blob A, has enough signal to allow for a meaningful fit.
    For blob A, we obtain an observed major axis of 0.64\arcsec$\pm$0.23\arcsec, an observed minor axis of 0.31\arcsec$\pm$0.6\arcsec, and a position angle of -75.8$\pm$14.9 degrees.
    The observed major axis is elongated approximately E-W, along the direction where the lens model predicts the greatest shear.
    The lens model applied in Sect.~\ref{sectlensprop} provided the estimated magnification for each blob along the major and minor axis.
    Correcting the sizes obtained from CO(5-4) fits in the $uv$ plane, we obtain a mean delensed size for RV4 of 0.40\arcsec$\pm$0.09~$\times$~0.26\arcsec$\pm$0.06, corresponding to a physical size of 2.64$\pm$0.58\,kpc~$\times$~1.71$\pm$0.38\,kpc.
    RV4 is thus relatively compact.

\begin{figure}%[!h] 
  	    \includegraphics[width=\columnwidth]{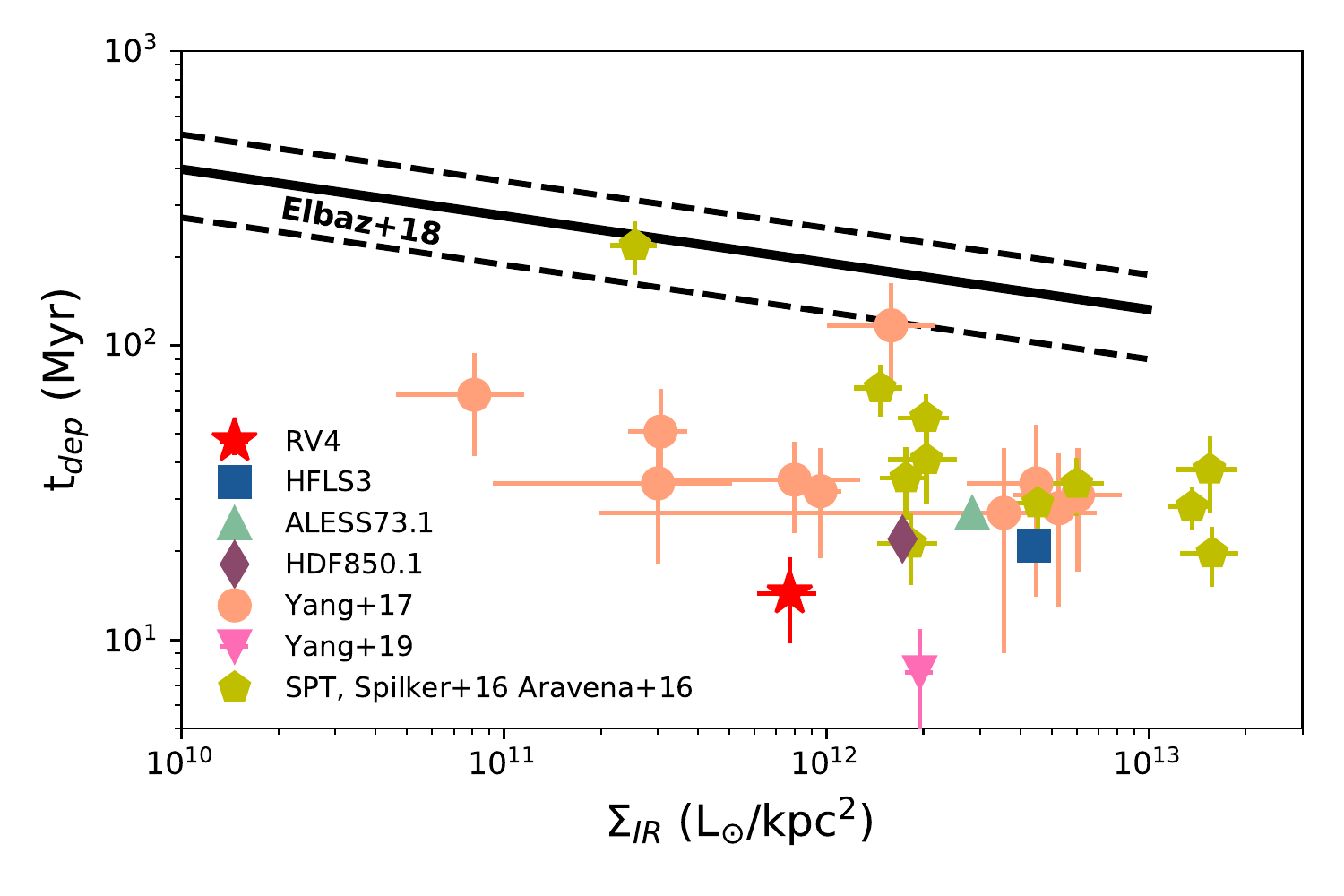}
  	    \caption{ \label{tdepl} Depletion time as a function of IR surface brightness. The position of RV4 is indicated by the red cross. The depletion time is the one computed from the gas mass estimated from [CI] (Sect.\,\ref{gas}). The blue square is HFLS3 \citep{Riechers13} at $z=$6.34, the green triangle is ALESS73.1 at $z=$4.76 \citep{deBreuck14}, the purple diamond is HDF850.1 \citep{Neri14} at $z=5.19$, and the pink downward triangle is the source of \citep{Yang19} at $z=$6.5.The orange dots are the strongly lensed sample of SMG of \cite{Yang17} spanning a redshift range between 2.09 and 4.24. The yellow pentagons are SPT sources from \cite{Spilker16} and \cite{Aravena16}. The relation obtained for a sample of MS galaxies at $z=2$ by \cite{Elbaz18} is shown in black solid and dashed lines in the $\Sigma_{IR}$ range probed by their data.
	}
    \end{figure}

    Although lensing applies a shear to RV4's morphology, the IR surface brightness is conserved.
    Adopting the lens model that best reproduces the positions of RV4's blobs as well as their flux ratios, we distribute the IR luminosity among the three blobs taking into account their respective modeled magnifications.
    Using the continuum sizes derived from the NOEMA data, we obtain surface brightnesses of $\Sigma_{IR}=9.5\pm2.0\times10^{11}\,$L$_{\odot}$\,kpc$^{-2}$, $7.1\pm1.5\times10^{11}\,$L$_{\odot}$\,kpc$^{-2}$, and $6.5\pm1.4\times10^{11}\,$L$_{\odot}$\,kpc$^{-2}$ for blobs A, B, and C, respectively.
    These values are consistent within the error bars and provide an average $\Sigma_{IR}$ of $7.7\pm1.6\times10^{11}\,$L$_{\odot}$\,kpc$^{-2}$ for RV4.
    In Fig.~\ref{tdepl}, we compare RV4's $\Sigma_{IR}$ to those of other SMGs found in the literature.
 
    RV4 has a notably low depletion time despite a $\Sigma_{IR}$ only slightly smaller than the sample median.
    Different assumptions could put RV4 closer to the median along both axis: the system's depletion times obtained from dust mass and gas-to-dust ratios are longer than the ones obtained from both [CI] and CO(5-4).
    Furthermore, if an obscured AGN is contributing to the L$_{IR}$ then the depletion time for RV4 would be higher.
    However, we do not expect these effects could erase all of the current shortfall relative to the \cite{Elbaz18} MS relation, although the evolution of such relation with redshift is not yet understood.
    We also note that our method uses the size derived from 3\,mm continuum emission which traces the relatively cold dust component ($\sim$520\microns, rest frame) whereas the bulk of the IR emission comes from star-forming regions.
    Such an offset could have an impact on our derived $\Sigma_{IR}$ although it is difficult to quantify with the data we have in hand.

\section{\label{shortdepl}Interpreting the short depletion time}
The depletion times estimated through different assumptions ([CI], CO, and dust) yield rather low values 14.4$\pm$4.7, 17.2$\pm$2.2, and 32.9.0$\pm$18.7\,Myr, respectively, strongly suggesting that RV4 is experiencing a starburst event.
As shown in Fig.~\ref{lci_lco}, the $L_{CI}/L_{IR}$ ratio is very low compared to a literature sample and seems to indicate a high intensity of the radiation field, close to those measured in quasars.

\begin{figure}%[!h] 
  \includegraphics[width=\columnwidth]{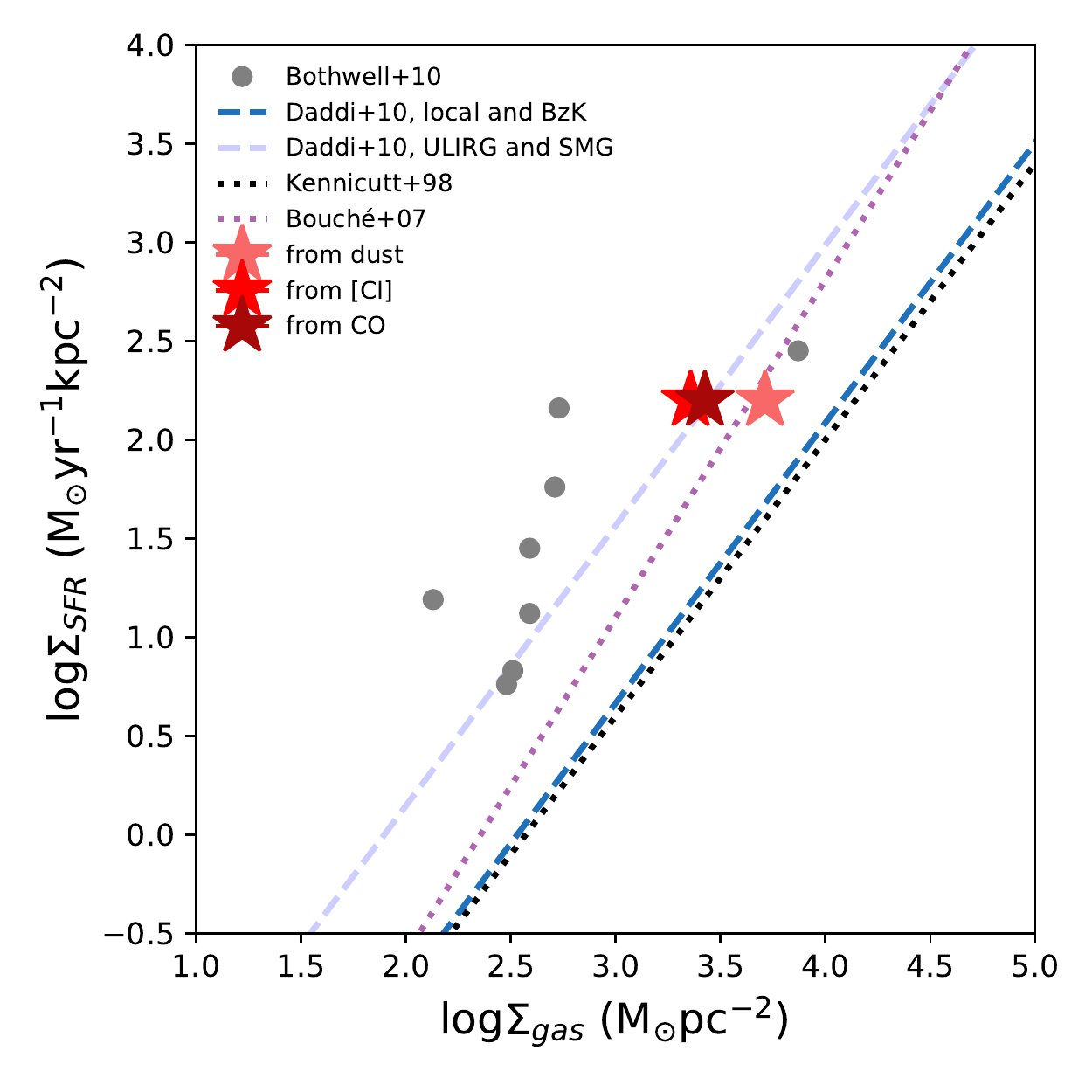}
  \caption{ \label{KS} Star formation rate surface density as a function of gas surface density. The position of RV4 is marker by the red stars. The different colors correspond to the method used to derive the gas mass: light red using the dust mass, red using [CI], and dark red using CO. The grey points are the SPT data from \cite{Bothwell10} which properties are similar to RV4's. The black dotted line is the \cite{Kennicutt98} relation and the purple dotted line the updated relation from \cite{Bouche07}. The blue and light blue dashed lines are the relation of \cite{Daddi10b} for BzK and SMG galaxies, respectively. 
}
\end{figure}

In Fig.\ref{KS}, we place RV4 in a Kennicutt-Schmidt diagram \citep{Kennicutt98} along with SPT sources from \cite{Bothwell10} whose properties are close to those of RV4.
The $\Sigma_{gas}$ determined from the CO and [CI] lines places RV4 on the relation of \cite{Daddi10a} for ULIRGs and SMGs. 
The $\Sigma_{gas}$ determined from M$_{dust}$ is larger but still above the relation of \cite{Daddi10a} for BzK galaxies but right on the universal relation determined by \cite{Bouche07}.
The position of RV4 in this diagram confirms the starbursting phase suggested by the very short depletion time.

Even considering a starburst, the typical depletion timescale of a starburst being 100\,Myr \citep[e.g.,][]{Bethermin15,Aravena16}, RV4's $t_{depl}$ is at least a factor 2 lower.
A first explanation to this short depletion time could be that RV4 has an unusually high star formation efficiency, defined as the star formation rate divided by the gas mass.  
However, such high star formation efficiencies are hard to explain theoretically and it has been shown in simulations that major mergers are not that efficient in producing high bursts of star formation in massive gas-rich galaxies \citep{Fensch17}.
A second possibility would be that RV4 encounters a ``recent'' and rapid decrease of the SFR.
In this case, the L$_{IR}$ would trace the SFR over a typical timescale of 100\,Myr \citep{KennicuttEvans12} while the gas mass is probed by emission lines which trace the current gas content.
If the SFR and the gas content are both rapidly decreasing, the L$_{IR}$ would overestimate the instantaneous SFR and thus bias the depletion time toward lower value.
Therefore, if the decrease of star formation is very fast the actual depletion time would be closer to typical starburst values.

If we assume that RV4 lies at least a factor of three above the MS, the stellar mass will be at most 1.7$\times$10$^{11}$\,M$_{\odot}$.
Considering the three estimates of the gas mass derived in Sect.\ref{gas}, corrected for magnification, we find gas fractions larger than 16\%, 18\%, and 30\% considering the [CI], CO, and dust based estimates, respectively.
These values are not constraining enough for SMGs for which gas fractions are $\sim$50\% \citep[e.g.][]{Daddi10b,Tacconi10}; therefore we cannot disentangle between the two possible explanations for the short depletion time.
Observations of RV4 with the JWST to probe its stellar content will be key to disentangle between the two scenarios.

%=================================================================================
\section{\label{conclusion}Conclusion}
We have serendipitously discovered a bright submillimeter galaxy in the field of view of the \textit{Herschel} imaging of the nearby galaxy IC\,3521.
The rising SPIRE colors and high flux densities indicate a high-z galaxy with high IR luminosity.
We have gathered ancillary data to shed light on the nature of this source and reached the following conclusions.

The redshift of RV4 is 4.724 as constrained from IRAM/EMIR and NOEMA data. The combination of \textit{Herschel} and IRAM/NIKA data constrains the IR SED of RV4 resulting in a total IR luminosity of 1.06$\pm$0.6$\times$10$^{14}$L$_{\odot}$. However, the data in hand do not allow us to exclude the presence of a dust enshrouded AGN that could be contributing to this IR luminosity.  

The \textit{Herschel} emission splits into three blobs in VLA at 10\,cm imaging. The IRAM/NOEMA detection of the CO(5-4) line of each blob confirms that the three sources are at the same redshift with CO(5-4) lines having the same width. Combined with the extreme L$_{IR}$, we conclude that RV4 is lensed. 

In \textit{Spitzer}/IRAC and deep CFHT data, two sources acting as a gravitational lens are detected at the center of the virtual arc formed by the three VLA/NOEMA blobs. These two sources, High-z Lens 1 West and East (HL1-W and HL1-E), are located at $z=$1.48 by XSHOOTER data, indicating that they are probably merging. They constitute one of the most distant gravitational lenses found to date in the literature for strong lensing systems. The Einstein radius of the lensing system is 2.2\arcsec$\pm$0.2, as determined from the positions of the three VLA/NOEMA blobs. HL1 is a very peculiar lens with a large Einstein radius combined with a high redshift. We derive a lens model and find that a single halo best reproduces both the positions and the flux ratios of the VLA/NOEMA blobs. This model yields a total magnification of 8.2$\pm$2.5 and thus an intrinsic L$_{IR}$ of 1.29$\times$10$^{13}$\,L$_{\odot}$ for RV4. A detailed model of the system will be presented in a companion paper (Ciesla et al., in prep).

The SED modelling of HL1-W and HL1-E yields stellar masses of 2.4$\pm$0.4 and 3.3$\pm$0.6$\times$10$^{11}$M$_{\odot}$, respectively, and SFR compatible with the two galaxies being passive. The absence of H$\alpha$ emission in HL1-W confirms the low SFR obtained from SED modeling. For HL1-E, a strong H$\alpha$ line is detected together with a strong and broad [NII] line, indicating the possible presence of an AGN. A detailed analysis of the HL1 system will be provided in the companion paper (Ciesla et al., in prep) as well.

From the IRAM/EMIR and NOEMA data, we have measured the [CI], CO(4-3), and CO(5-4) apparent luminosities. On the one hand RV4 shows a relatively low L$_{ [CI]}$ to L$_{IR}$ ratio, which can be interpreted as a high star formation efficiency compared to other high-z samples of galaxies, and/or with a hard radiation field. On the other hand, the L$_{ [CI]}$ to L$_{CO(4-3)}$ ratio of RV4 is relatively high, indicating a lower gas density. Furthermore, the L$_{CO(5-4)}$ to L$_{IR}$ is only half of the value expected for normal galaxies, indicating a lack of gas. 

We estimate the gas mass of RV4 from different tracers and obtain between 2.7$\pm$0.7$\times$10$^{11}$\,M$_{\odot}$ and 6.0$\pm$3.0$\times$10$^{11}$\,M$_{\odot}$, not corrected for magnification. These numbers yield depletion times between 14.4$\pm$4.7 and 32.9$\pm$18.7\,Myr. Such short values can be explained by either very high star formation efficiency or by a rapid and recent decrease of star formation. In the former scenario, RV4 would be an intense starburst, difficult to understand from the results of major mergers simulations of gas rich high-$z$ galaxies. In the second scenario, the L$_{IR}$ probing the star formation activity on a timescale of $\sim$100\,Myr and the gas content following the instantaneous SFR, if the star formation activity of RV4 is rapidly decreasing it would bias $t_{depl}$ toward low values. JWST will definitely be needed to determine the M$_{gas}$ to M$_*$ ratio of RV4 and investigate the possible low gas content of this SMG.

%=================================================================================
\begin{acknowledgements}
Based on observations carried out under project number P319809 with the
IRAM NOEMA Interferometer and project numbers 234-14 and D07-15 with the 30-meter telescope. %D285879
IRAM is supported by INSU/CNRS (France), MPG (Germany) and IGN (Spain). We would like to thank the IRAM staff for their support during the NIKA campaign.
MF is supported by a Royal Society - Science Foundation Ireland University research fellowship.
RJA was supported by FONDECYT grant number 1191124.
MB acknowledges the FONDECYT regular grant 1170618.
AJB acknowledges support from the National Science Foundation via grant AST-0955810.
CY acknowledges support from an ESO Fellowship.
We thank the referee for his/her comments that helped improving this paper.
\end{acknowledgements}

%%=================================================================================

\bibliographystyle{aa}
\bibliography{pcigale_bizarre_submm}

\end{document}